# Betrayal, Distrust, and Rationality: Smart Counter-Collusion Contracts for Verifiable Cloud Computing[*]

Changyu Dong[†]
Newcastle University
Newcastle Upon Tyne, UK
changyu.dong@newcastle.ac.uk

Yilei Wang
Newcastle University
Newcastle Upon Tyne, UK
yilei.wang@newcastle.ac.uk

Amjad Aldweesh
Newcastle University
Newcastle Upon Tyne, UK
a.y.a.aldweesh2@newcastle.ac.uk

Patrick McCorry[‡]
University College London
London, UK
p.mccorry@ucl.ac.uk

Aad van Moorsel
Newcastle University
Newcastle Upon Tyne, UK
aad.vanmoorsel@newcastle.ac.uk

## ABSTRACT
Cloud computing has become an irreversible trend. Together comes the pressing need for verifiability, to assure the client the correctness of computation outsourced to the cloud. Existing verifiable computation techniques all have a high overhead, thus if being deployed in the clouds, would render cloud computing more expensive than the on-premises counterpart. To achieve verifiability at a reasonable cost, we leverage game theory and propose a smart contract based solution. In a nutshell, a client lets two clouds compute the same task, and uses smart contracts to stimulate tension, betrayal and distrust between the clouds, so that rational clouds will not collude and cheat. In the absence of collusion, verification of correctness can be done easily by crosschecking the results from the two clouds. We provide a formal analysis of the games induced by the contracts, and prove that the contracts will be effective under certain reasonable assumptions. By resorting to game theory and smart contracts, we are able to avoid heavy cryptographic protocols. The client only needs to pay two clouds to compute in the clear, and a small transaction fee to use the smart contracts. We also conducted a feasibility study that involves implementing the contracts in Solidity and running them on the official Ethereum network.

## KEYWORDS
Verifiable Computing; Smart Contract; Game Theory; Collusion; Trust

## 1 INTRODUCTION
Cloud computing has reached critical mass and become indispensable to businesses. A 2016 report [42] found that 95% of the organizations surveyed are running applications or experimenting with the cloud. Data from Synergy Research Group [45] showed that the worldwide cloud computing market reached $148 billion in 2016, having grown by 25% on an annual basis. Gartner predicted more than $1 trillion in IT spending will be directly or indirectly impacted by the transition to cloud computing by 2020 [17].

In the context where organizations embrace clouds and reap clear business benefits, **verifiability** becomes a critical requirement for cloud computing. Cloud computing is a service provided by an external party, thus it is difficult for the client to fully trust the cloud provider. Should the cloud return a wrong result for a mission-critical task, the consequence would be disastrous. To exercise due diligence and gain greater confidence in computation outsourced to the cloud, clients need to be able to verify the correctness of the results returned.

Roughly, existing solutions for verifying outsourced computation are based on either cryptography or replication (see Section 9). Typically in the **cryptography-based** approach, the client outsources a task to a single cloud server. The cloud returns the computation result and proves to the client that the result was computed correctly. Cryptography ensures the client will reject with a high probability if the result is incorrect. In the **replication-based** approach, the client gives the same task to multiple clouds and the clouds compute the task independently. The client then collects and crosschecks the results. As long as the number of faulty servers is below a threshold, the correctness of result can be verified using a consensus protocol.

**A Cost Analysis** Existing verifiable computation techniques are not quite economically sound. The biggest motivation for businesses to adopt cloud computing is perhaps cost saving. For example, we used the Amazon AWS Total Cost of Ownership Calculator [3] on a few typical settings, and found that by moving their on-premises IT infrastructures to AWS, companies could save 50% to 69% of the cost (See Appendix A). The saving is large, however is not large enough to sustain existing verifiable computation techniques. Cloud computing is based on the pay-per-use paradigm and the clients are charged for the resources they use. Using the cryptography-based approach to verify a task in the cloud means that the client has to pay for the overhead imposed by the cryptographic algorithms/protocols. The typical overhead is $10^3$ - $10^9$ times higher than computing the task itself [51] and would translate to a prohibitively high financial cost to the client. The replication-based approach usually computes the task in the clear and the overheads mainly come from employing multiple replicas. Usually at least 3 replicas are required[1], which means the total cost to the client is at least tripled. From the cost saving figures showed earlier, it is clear that using 3 or more clouds for verification is very likely to cost more than simply using on-premises IT infrastructures.

---

[1]Except for [11], which uses a minimal of 2 replicas. However the protocol introduces an overhead that is about 10 - 20 times of the computation being verified.

**Problem Statement** In summary, we want verifiable cloud computing at a competitively low cost. The clients should be able to get a strong correctness guarantee of the computation in the clouds, and pay similar or less than what they pay when using on-premises IT infrastructures. To accomplish this, we opt for the replication based approach because it is much closer to being practical. It is clear from the analysis above that to make the financial cost of the solution competitive compared to on-premises IT, the client should pay no more than 2 replicas for the computation and should minimize other overheads. The biggest challenge of using only 2 replicas is *collusion*. If the two clouds coordinate and output the same wrong result, the client might accept the wrong result without even realizing it. It becomes even more challenging when heavy-weight cryptographic protocols have to be avoided in order to reduce the overhead to an acceptable level. To this end, we resort to game theory and a new financial instrument, namely *smart contracts*, for tackling the problems.

**The Idea** Rather than forbidding or preventing collusions through technical means such as cryptography, we work towards undermining, through economic means, the foundation that collusion is grounded on. This should not be surprising since collusion is a topic studied in economics for many years. Three insights from economists establish the premise of our work:

- Collusion occurs "whenever it is more profitable to all of the participants than their feasible alternatives" [46]. Since collusion is often driven by economic incentives, imposing high fines on collusion has become a major instrument for preventing collusions in the real world. The fines make collusion a less profitable choice than not colluding, thus offset the motivation for collusion.
- Colluding parties have their own interests, and this is a source of tension between them [34]. Colluding parties are not a single corporate entity. More interestingly, they are often competitors who collude in order to gain extra profit. Nevertheless, each party is responsible to its own and acts in its own interest. Under suitable conditions, collusion can dissolve and competition can resume.
- The most pressing problem for the colluding parties is how to prevent cheating. This is a natural consequence of pursuing self-interest, i.e. parties act in their own interest and try to maximize their own profit. In fact, "the central difficulty of collusion is that it is often profitable for firms to secretly deviate from the collusive agreement" [34].

Our key idea is to sabotage collusion by using smart contracts. Here smart contracts materialize self-enforcing agreements and payments that serve multiple purposes: (1) To weaken the incentive for collusion by taking a deposit from the clouds as security for the delivery of the correct result. The clouds will be penalized by losing their deposit should they deliver a wrong result. (2) To create an incentive for correct computation by redistributing the fine to the honest cloud as a reward. (3) To create distrust between the colluders by incentivizing them to betray their partner in the collusion coalition. On the whole, we intend to make collusion a less favorable choice and make it much harder for potential colluding parties to trust each other, so that rational parties will stay away from collusion because it is unprofitable and too risky.

**Contributions** Based on the idea above, we designed two smart contracts (the Prisoner's contract and the Traitor's contract) to be used in scenarios where a client outsources a computation task to two clouds and cross-checks the results from the two clouds. With moderate and reasonable assumptions, the contracts guarantee that the two clouds, if they are rational, will behave honestly even though they have the opportunity to collude together and cheat. We conducted detailed game theoretical analysis of the contracts. We proved that for the two clouds, both being honest and not colluding is the unique sequential equilibrium (a stronger form of Nash equilibrium) of the game. We also show feasibility of the contracts by building them for the Ethereum network. We created the contracts using Solidity and executed them on the official Ethereum network. We provide a breakdown of financial and computational overheads for our contracts. Our figures show that the total transaction cost for executing each contract is below $1.

The **Prisoner's contract** is to be signed by a client and two clouds. The name comes from the fact that the contract induces a game similar to the famous Prisoner's Dilemma game between the two clouds. At a high level, the contract says that the client will pay the two clouds to compute a task, but to get the job, each cloud has to pay a deposit. The honest cloud will get its deposit back later, the cheating cloud will lose its deposit (if cheating is detected). Moreover, if one cloud cheats and one cloud is honest, the cheating cloud's deposit goes to the honest cloud as a bonus (after deducting certain necessary costs). Similar to in the Prisoner's Dilemma game, although it seems both clouds gain most by colluding with each other, both clouds eventually end up being honest. This is because they know the other will act in its own interest, which means they will deviate from collusion for a higher payoff.

The problem with the Prisoner's contract is that it only works if the two clouds cannot make credible and enforceable promises. This is not true especially with the help of smart contracts. We demonstrate this by the **Colluder's Contract**, which is a secret smart contract between the two clouds. In the contract, the cloud who initiates the collusion coalition agrees to pay a bribe to incentivize the other cloud to collude. More importantly, both clouds make a commitment by paying a deposit which will be taken if they do not follow the collusion strategy. The contract totally changes the game: when the deposit is high enough to offset the benefit a cloud can gain by betraying the other, betrayal is no longer more profitable and collusion becomes the best strategy for both clouds.

To bust this form of more robust collusion coalition policed by collusion agreements such as the Colluder's contract, we designed the **Traitor's contract**. Intriguingly, the Traitor's contract works not by countering the collusion agreement directly, but by forgiving one (and only one) cloud who follows the collusion strategy. The aim of the Traitor's contract is not to incentivize the clouds to deviate from the collusion, but to encourage them to report the collusion to the client. By getting information about collusion, the client can further investigate the case and punish the cheating cloud. By following the collusion strategy, the reporting cloud avoids the punishment imposed by the collusion agreement thus making the agreement useless. If the other cloud does cheat, the reporting cloud will get a reward, which makes reporting the most profitable strategy[2]. Overall, reporting is risk-free (the reporting cloud will not be punished by the Prisoner's contract and the Colluder's contract) and more profitable. The consequence is that both clouds know that if they try to initiate a collusion coalition, the other will collude but also report it to the client. This creates distrust between the clouds so that neither will want to initiate the collusion coalition, and they will stay honest to avoid being betrayed and punished.

The main cost of our smart contract based solution is the cost for employing two clouds to compute (in the clear) the same task. We assume that an offline Trusted Third Party (TTP) is available to resolve the dispute when an inconsistency or anomaly is detected. However, if the two clouds are rational, the TTP will never be involved. Even if in the unlikely cases the TTP is called upon, the cost for dispute resolution is borne by the faulty cloud, not the client. The implementation of the contract requires only a few (constant number) additional cryptographic operations that are very light. Our experiments on the official Ethereum network show that the transaction cost for using smart contract facilities is small.

## 2 PRELIMINARIES

In Section 2.1 and 2.2, we briefly review relevant concepts in game theory. The two sections are mostly based on [33, 35].

### 2.1 Games and Strategies

In this paper, we describe games in *extensive form with imperfect information*. In extensive form, a game is depicted as a game tree. The tree shows choices and information available to players when they are called to take an action, the order in which players make their moves, the outcomes of the game and the payoffs of the outcomes. Imperfect information means that the players may not know all the actions taken by the other players. Imperfect information is more realistic and allows a wider scope of analysis than perfect information, i.e. assuming players knows every move of the others. For example, simultaneous moves and deception can be captured by imperfect information. Formally, we have:

*Definition 2.1.* A **finite game in extensive form with imperfect information**, or **game** for short in this paper, is a tuple $\mathcal{G} = (\mathcal{N}, \mathcal{A}, \mathcal{H}, \mathcal{Z}, \chi, \rho, \sigma, u, \mathcal{I})$ where:

- $\mathcal{N}$ is a set of $n$ players.
- $\mathcal{A}$ is a single set of actions.
- $\mathcal{H}$ is a set of nonterminal choice nodes.
- $\mathcal{Z}$ is a set of terminal nodes, disjoint from $\mathcal{H}$.
- $\chi : \mathcal{H} \to 2^{\mathcal{A}}$ is the action function, which assigns to each choice node a set of possible actions.
- $\rho : \mathcal{H} \to \mathcal{N}$ is the player function, which assigns to each nonterminal node a player $i \in \mathcal{N}$ who chooses an action at that node.
- $\sigma : \mathcal{H} \times \mathcal{A} \to \mathcal{H} \cup \mathcal{Z}$ is the successor function, which maps a choice node and an action to a new choice node or terminal node such that for all $h_1, h_2 \in \mathcal{H}$ and $a_1, a_2 \in \mathcal{A}$, if $\sigma(h_1, a_1) = \sigma(h_2, a_2)$ then $h_1 = h_2$ and $a_1 = a_2$.
- $u = (u_1, ..., u_n)$ where $u_i : \mathcal{Z} \to \mathbb{R}$ is a real-valued utility function for player $i$ on the terminal nodes $\mathcal{Z}$.

[2]Reporting is most profitable only if collusion happens. The contract has clauses to punish a cloud that misreports a fabricated case.

- $\mathcal{I} = (\mathcal{I}_1, ..., \mathcal{I}_n)$ where $\mathcal{I}_i$ is an equivalent relation that partitions player $i$'s choice nodes $\{h \in \mathcal{H} : \rho(h) = i\}$ into $k_i$ information sets $\mathcal{I}_{i,1}, ..., \mathcal{I}_{i,k_i}$ with the property that $\chi(h) = \chi(h')$ and $\rho(h) = \rho(h')$ whenever there exists a $j$ for which $h \in \mathcal{I}_{i,j}$ and $h' \in \mathcal{I}_{i,j}$.

In the definition, $(\mathcal{N}, \mathcal{A}, \mathcal{H}, \mathcal{Z}, \chi, \rho, \sigma, u)$ captures the setting and rules of the game, and $\mathcal{I}$ captures the imperfection of information. An example game is shown in Figure 1. In the game tree, we use circles for choice nodes and rectangles for terminal nodes. Each node has a label $v_i$. In the game, there are two players $\mathcal{N} = \{P_1, P_2\}$. Actions available to players are $\mathcal{A} = \{L, M, R, \ell, r, x, y\}$. The choice nodes are $\mathcal{H} = \{v_0, v_2, v_3, v_4, v_5, v_6, v_7\}$ and the terminal nodes are $\mathcal{Z} = \{v_1, v_8, v_9, v_{11}, v_{12}, v_{13}, v_{14}, v_{15}\}$. The function $\chi$ assigns actions to choice nodes. Actions $\{L, M, R\}$ are assigned to $v_0$. The nodes $v_2$ and $v_3$ are both assigned $\{\ell, r\}$, and the nodes $v_4$ to $v_7$ all have the same actions $\{x, y\}$. The function $\rho$ assigns choice nodes to players. In the figure, we label the choice nodes with its player. In the game $P_1$ has $\{v_0, v_4, v_5, v_6, v_7\}$ and $P_2$ has $\{v_2, v_3\}$. The function $\sigma$ is captured in the tree structure by the parent-child relationship. After a player chooses an action, the game will move to the child node following the edge labeled with the action. The utility of outcomes for each player are displayed at the bottom under the leaf nodes. Information sets are represented as elongated dashed circles encompassing some nodes, unless the information set has only one node. So there are 3 information sets for $P_1$: $\mathcal{I}_{1,1} = \{v_0\}, \mathcal{I}_{1,2} = \{v_4, v_5\}, \mathcal{I}_{1,3} = \{v_6, v_7\}$, while only 1 information set for $P_2$: $\mathcal{I}_{2,1} = \{v_2, v_3\}$. A player cannot distinguish nodes in the same information set. For example, after $P_1$ has made the first move, $P_2$ does not know whether he is at $v_2$ or $v_3$ because he does not know whether $P_1$ chose $M$ or $R$.

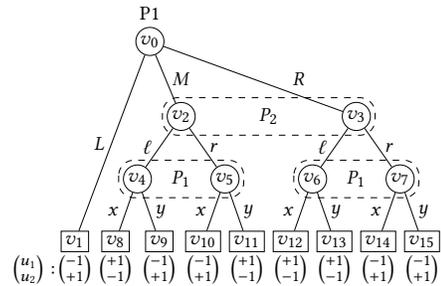

**Figure 1: An Example Game**

Strategies determine the action a player will take at any stage of a game. In this paper, we focus on *behavior strategies* which are more general than pure strategies and are equivalent to mixed strategies in our setting.

*Definition 2.2.* Let $\mathcal{G}$ be a game, a **behavior strategy** $s_i$ of player $i$ is a function that assigns each information set $\mathcal{I}_{i,j} \in \mathcal{I}_i$ a probability distribution over the actions in $\chi(\mathcal{I}_{i,j})$, with the property that each probability distribution is independent of the others. A **completely mixed behavior strategy** is a behavior strategy in which every action is assigned a positive probability. A **strategy profile** is a list of all players' strategies $s = (s_i)_{i \in \mathcal{N}}$. A strategy profile without player $i$'s strategy is defined as $s_{-i} = (s_1, ..., s_{i-1}, s_{i+1}, ..., s_n)$. We can also write $s = (s_i, s_{-i})$.

For example the game in Figure 1 can have a strategy profile $(s_1, s_2)$ where $s_1 = \left(\left[\frac{1}{3}(L), \frac{1}{3}(M), \frac{1}{3}(R)\right], \left[\frac{3}{4}(x), \frac{1}{4}(y)\right], \left[\frac{1}{2}(x), \frac{1}{2}(y)\right]\right)$, $s_2 = ([1(\ell), 0(r)])$. For $P_1$, the strategy $s_1$ says that to play all actions at information set $\mathcal{I}_{1,1}$ with a equal probability of $\frac{1}{3}$, to play $x$ with a probability of $\frac{3}{4}$ and $y$ with a probability of $\frac{1}{4}$ at information set $\mathcal{I}_{1,2}$, and to play $x$ and $y$ with a equal probability of $\frac{1}{2}$ at information set $\mathcal{I}_{1,3}$. For $P_2$ the strategy $s_2$ says that to play $\ell$ for sure and never play $r$ at information set $\mathcal{I}_{2,1}$.

## 2.2 Sequential Equilibrium

The most important solution concept in game theory is the Nash equilibrium. Informally, in a Nash equilibrium, every player's strategy is the best given the other players' strategies, and no one can do better by changing strategy if the others do not change their strategies. However Nash equilibria can be weak sometimes. In some Nash equilibria, it is possible that a player's strategy includes irrational actions (non-credible threats) that lead to a lower payoff for himself. There are several refinements of the Nash equilibrium that aim to exclude those implausible equilibria. A stringent and influential refinement is the *sequential equilibrium* [27]. Sequential equilibria exclude weak strategies by requiring a strategy to be sequentially rational, i.e. optimal not just in terms of the whole game but also at each information set. The sequential equilibrium can also be seen as a refinement of other popular refinements e.g. subgame perfect equilibrium and perfect-Bayes equilibrium.

A sequential equilibrium is comprised of a strategy profile and a belief system. With imperfect information, players have to make decisions under uncertainty. When the player is called to make a decision, he needs beliefs of where he is in the game tree. The belief system allows players to construct a strategy that is optimal at every point in the tree.

*Definition 2.3.* In a game $\mathcal{G}$, a **belief system** $\beta = (\beta_i)_{i \in \mathcal{N}}$ is the following: for each player $i$, $\beta_i$ assigns each information set $\mathcal{I}_{i,j} \in \mathcal{I}_i$ a probability distribution over the nodes in $\mathcal{I}_{i,j}$. For each node $h \in \mathcal{I}_{i,j}$, the belief $\beta_i(h) = Pr[h|\mathcal{I}_{i,j}]$, i.e. the probability that player $i$ is at $h$ given that he is at $\mathcal{I}_{i,j}$.

*Definition 2.4.* In a game $\mathcal{G}$, the player $i$'s **expected payoff at** $h$, given the play of the game is at node $h$ when the players implement the strategy profile $s$, is the sum of the utility of each terminal nodes, weighted by the probability of reaching the node:
$$u_i(s; h) = \sum_{z \in \mathcal{Z}} Pr[z|(s, h)] \cdot u_i(z)$$

The player $i$'s **expected payoff at** $\mathcal{I}_{i,j}$ is the sum of expected payoff at each $h \in \mathcal{I}_{i,j}$, weighted by the belief $\beta_i(h)$:
$$u_i(s; \mathcal{I}_{i,j}, \beta) = \sum_{h \in \mathcal{I}_{i,j}} \beta_i(h) \cdot u_i(s; h)$$

*Definition 2.5.* An **assessment** is a pair $(s, \beta)$ in which $s$ is a behavior strategy profile and $\beta$ is a belief system.

*Definition 2.6.* Let $\mathcal{G}$ be a game, $(s, \beta)$ be an assessment, the strategy profile $s = (s_i, s_{-i})$ is called **rational** at information set $\mathcal{I}_{i,j}$, relative to $\beta$, if for each behavior strategy $s'_i \neq s_i$ of player $i$:
$$u_i(s; \mathcal{I}_{i,j}, \beta) \geq u_i((s'_i, s_{-i}); \mathcal{I}_{i,j}, \beta)$$

The assessment is called **sequentially rational** if for each player $i$ and each information set $\mathcal{I}_{i,j} \in \mathcal{I}$, the strategy profile $s$ is rational at $\mathcal{I}_{i,j}$ relative to $\beta$.

*Definition 2.7.* An assessment $(s, \beta)$ is said to be **consistent** if there exists a sequence of fully mixed behavior strategy profiles $(s^k)_{k \in \mathbb{N}}$ satisfying the following conditions:
(1) The profile $(s^k)_{k \in \mathbb{N}}$ converges to $s$, i.e. $\lim_{k \to \infty} (s^k) \to s$;
(2) The sequence of beliefs $(\beta^k)_{k \in \mathbb{N}}$ induced by $(s^k)_{k \in \mathbb{N}}$ (by Bayes' rule) converges to the belief system $\beta$, i.e. $\lim_{k \to \infty} (\beta^k) \to \beta$;

*Definition 2.8.* An assessment $(s, \beta)$ is called a **sequential equilibrium** if it is sequentially rational and consistent.

Given a sequential equilibrium, the expected utility of each player *at any point* of the game is the highest given the strategy profile and his beliefs. Therefore a rational player will not deviate from the equilibrium. Consistent (Definition 2.7) ensures that the beliefs match the strategy profile by requiring the beliefs to be derivable from the strategy profile by applying Bayes' rule. The two conditions in the Definition 2.7 ensure this is true even with the unreachable information sets that are not on the equilibrium path.

## 2.3 Smart Contracts

**Cryptocurrencies** have gained great popularity recently. The idea of cryptocurrencies is grounded on a decentralized network of peers that provides the infrastructure to maintain a public ledger, which stores all transactions of the network. The ledger is stored in the form of a blockchain whose state is agreed by the peers through a consensus protocol. As the name indicates, cryptocurrencies use cryptography to secure transactions and to control the creation of additional units of the currency. **Smart contracts** are machinery built on top of cryptocurrencies to allow defining and executing contracts on the blockchain. In the simplest terms, a smart contract is a piece of computer program stored and running on the blockchain. The program code captures the logic of contractual clauses between parties. The execution of the code is triggered by events e.g. transactions added to the blockchain. The code is executed by the consensus peers and the correctness of execution is guaranteed by the consensus protocol of the blockchain. Ideally, we can think smart contracts as being executed by a trusted global machine that will faithfully execute every instruction.

Ethereum [14] is perhaps the most prominent example of cryptocurrencies that support smart contracts. In Ethereum, the currency is called **ether**. Ethereum contracts can be written in various expressive scripting languages such as **Solidity**. Ethers are held in and can be transferred between accounts. There are two types of **accounts**: externally owned accounts and contract accounts. An externally owned account is associated with a unique public-private key pair, owned by someone and has an ether balance in it. The owner has the private key that can be used to sign transactions from this account. Contract accounts do not have an associated private key. It maintains an ether balance and stores the code of a contract that decides the flow of the ethers in the account. A **transaction** in Ethereum is an instruction that is constructed and cryptographically signed by an externally owned account owner. Each transaction has two address fields that specify the sender and the receiver. One can initiate a contract by creating a transaction in which the receiver is a new contract account address and the `data` field contains the contract code. A transaction can also be used as a message to invoke a function in a contract. In this case the receiver's address is the contract account storing the contract code and the function to be invoked along with arguments is specified in the `data` field of the transaction. The behavior of a contract is purely decided by the execution of its code. A transaction also includes some "**gas**" and a "**gas price**" [52]. Executing a transaction will consume gas and the amount of gas consumed is converted into ether using the gas price, the ether is charged to the sender's account as the transaction fee.

## 3 ADVERSARY MODEL AND ASSUMPTIONS

Following the convention of the verifiable computation literature, in this paper, we consider only the integrity of the computation but not the confidentiality. We consider an *honest* client who outsources a computation task and pays two clouds to compute the result. For the client, the goal is to get the correct result while minimizing the cost. The clouds are unreliable and can return wrong results for outsourced computation tasks. Note that in this paper, we do not distinguish intentional and unintentional faults because it is difficult to collect evidence. If trusted auditing services are available to provide proper evidence then these two types of faults can be treated differently. We assume the clouds are physically isolated and model each cloud as an individual *rational adversary*. Rational means that a party always acts in a way that maximizes its payoff, and is capable of thinking through all possible outcomes and choosing strategies which will result in the best possible outcome. Compared to assuming a malicious adversary who will act arbitrarily, rational is more realistic when modeling corporate behavior of the clouds. Indeed, a cloud provider is more likely to cut corners in order to maximize its profit than maliciously attack the client with no reason. On the other hand, rational adversaries are weaker than malicious adversaries because rationality precludes certain strategies. There is a trade-off between the level of security guarantee and costs. In the case that adversaries may behave irrational, cryptography-based approaches could be used to ensure verifiability.

We assume incorrect computation costs less (e.g. by skipping part or all of the computation), so the clouds are motivated to cheat. For simplicity, we assume a cloud can come up with an incorrect but plausible answer (cannot be easily proved to be wrong) at no cost. In reality this is not free. However, assuming such an answer can be picked with no cost guarantees that the lower bound of deposits we derive later is always valid because the cheating cloud loses strictly more if the cost of picking such an answer is more than 0. We view collusion as coordinated actions that follows from a mutual agreement between the adversaries. In reality, even if parties collude, they still retain their separate judgement and act in their own interests. Therefore modeling each cloud as an individual adversary is more realistic than as a monolithic adversary who corrupts and controls multiple clouds. We assume the adversaries are computationally bounded so all cryptographic primitives we need to use remain secure.

We assume there exists one or more cryptocurrencies that support smart contracts. Most smart contracts platforms are experimental now but there has been much effort to bring them into the real world. We assume the currency in these systems carries a certain amount of monetary value and is accepted by all parties under consideration as a medium of exchange. We assume the value of the currency is stable during the whole lifetime of the contract (and contracts derived from it). We assume the cryptocurrencies are secure and the smart contracts are executed faithfully.

We assume the existence of a trusted third party (TTP), who is offline most of the time but can be called upon to recompute the task and resolve any disputes. We stress that if the clouds are rational, then the TTP would never be involved. The very existence of such a TTP provides a deterrence power which the adversaries have to take into account when making decisions. Even without taking actions, the TTP is a tangible threat to the adversaries and will have a controlling influence over them. The idea is similar to some strategic concepts in modern warfare and politics, e.g. "fleet in being" and "nuclear deference".

We also assume the following:

- The task to be computed is deterministic or can be reduced to being deterministic, e.g. by providing a seed and using a pseudo-random generator for the random choices if the task is probabilistic. This is a common requirement in replication-based verifiable computation. We also assume the probability of guessing the correct result is small (e.g. by using inner state hash [6]).
- The task to be computed is not time-critical. We rely on the smart contract network to enforce the contracts, which may have large latency. The latency greatly depends on the status and parameters of the smart contract network and we will unlikely to get any guarantee for time-critical tasks.
- The parties can communicate freely and choose strategically what to say and what not to say. They communicate through reliable authenticated public or private channels.
- For simplicity, we assume all clouds have an equal cost for computing the same task and the cost is public. In reality this assumption does not always hold. Nevertheless, cost is not a decisive factor in the game. Therefore, assuming equal cost does not affect the analysis.
- The client is resource-constrained, i.e. it is not capable of recomputing the task to verify the result. In this case, proving faults of cloud can be difficult for the client and the TTP is necessary. We also assume the client is lazy, i.e. it will not ask the TTP to recompute the task unless there is clear evidence that this is necessary.
- Funds only flow among the parties under consideration, not to/from external parties. For example we do not consider fines imposed by legal systems or bribes offered by the client's rival in exchange for the clouds to output a wrong result. In general, if the cloud can gain additional benefits, one solution could be to increase the deposit. When the increment of deposit is large enough and surpasses the benefit, the cloud will behave honestly because otherwise the payoff will be worse than behaving honestly.
- Parties are risk neutral. For other risk profiles (risk seeking or risk aversion), the utility function can be adjusted to the risk profile and the equilibria still hold by choosing the deposits according to the risk profile.

## 4 MONETARY VARIABLES

Below are the monetary variables we will use in the contracts (listed in alphabetic order). They are all non-negative.

- $b$: the bribe paid by the ringleader of the collusion to the other cloud in the collusion agreement (the Colluder's contract).
- $c$: the cloud's cost for computing the task.
- $ch$: the fee to invoke the TTP for recomputing a task and resolving disputes.
- $d$: the deposit a cloud needs to pay to the client in order to get the job.
- $t$: the deposit the colluding parties need to pay in the collusion agreement (the Colluder's contract).
- $w$: the amount that the client agrees to pay to a cloud for computing the task.
- $z$: shorthand for $w - c + d - ch$

The following relations hold for obvious reasons:

- $w \geq c$: the clouds do not accept under-paid jobs.
- $ch > 2w$: otherwise there is no need to use the clouds, the client just uses the TTP for the computation. Note that $ch$ will be paid by the cheating cloud. An honest client pays strictly no more than hiring two clouds (plus the mere transaction cost).

The following relations needs to hold when setting the contracts in order for the desirable equilibria to hold. The parameter $d$ can be set by the client in the Prisoner's contract, $b$ and $t$ can be set by the clouds in the Colluder's contract (see explanations in later sections):

- $d > c + ch$
- $b < c$
- $t > z + d - b$

## 5 THE PRISONER'S CONTRACT

### 5.1 The Contract

The Prisoner's contract is an outsourcing contract signed between a client and two clouds. At a high level, it tries to incentivize correct computation by asking the clouds to pay a deposit upfront. If a cloud behaves honestly, the deposit will be refunded; if a cloud cheats (and is detected), the deposit will be taken by the client. Moreover, in the case where one cloud is honest and one cheats, the honest cloud gets an additional reward that comes from the deposit of the cheating cloud. The intuition is to create a Prisoner's dilemma between the clouds: although collusion leads to a higher payoff than both behaving honestly, there is an even higher payoff if one can lure the other into cheating while being honest itself. Once both clouds understand this, they know collusion is not stable because the other cloud will always try to deviate from it. Any attempts (without a credible and enforceable promise) to persuade the other to collude will be deemed to be a trap and thus will not be successful. The contract is presented below and more comments will follow afterwards.

(1) The contract should be signed between a client (CLT) and two clouds ($C_1, C_2$). Should there be any dispute, the dispute will be resolved by a trusted third party TTP.
(2) $C_1, C_2$ agree to compute a function $f()$ on an input $x$. Both $f()$ and $x$ are chosen by CLT.
(3) The parties agree on deadlines $T_1 < T_2 < T_3$.
(4) CLT agrees to pay $w$ to each cloud for the correct and timely computation of $f(x)$.
(5) As a condition, each of $C_1, C_2$ must pay a deposit of amount $d$ when signing the contract. The deposit will be held by the smart contract.
(6) $C_1, C_2$ must pay the deposit before $T_1$. If any $C_i$ fails to do so, the contract will terminate and any deposit paid will be refunded.
(7) $C_1, C_2$ must deliver the computation result $f(x)$ before $T_2$.
(8) Upon receiving the computation result from both $C_1, C_2$, or when the deadline $T_2$ has passed, CLT should do the following:
   (a) If both $C_1, C_2$ failed to deliver the result, their deposits will be taken in full by CLT;
   (b) If both $C_1, C_2$ delivered the result, and the results are equal, then after verifying the results, CLT must pay the agreed amount $w$ and refund the deposit $d$ to each $C_i$;
   (c) Otherwise CLT will raise a dispute to TTP.
(9) Upon receiving a dispute raised by CLT, TTP computes $f(x)$. Let $y_t, y_1, y_2$ be the results computed by TTP, $C_1, C_2$ respectively. Then the cheating party can be decided by the following rule:
   (a) For each $C_i$, if $C_i$ failed to deliver the result, $C_i$ cheated;
   (b) For each $y_i$ ($i \in \{1, 2\}$) delivered before the deadline, if $y_i \neq y_t$, $C_i$ cheated;
   TTP communicates the decision to CLT as well as to $C_1, C_2$.
(10) Upon receiving TTP's decision, the dispute is resolved as follows:
   (a) If none of $C_1, C_2$ cheated, CLT must pay the agreed amount $w$ and refund the deposit $d$ to each $C_i$, and pay the fee for resolving the dispute $ch$ to TTP.
   (b) If both $C_1, C_2$ cheated, their deposits will be taken in full by CLT, and CLT pays the fee $ch$ to TTP.
   (c) If only one of $C_1, C_2$ cheated, then (1) the deposit of the cheating cloud will be taken in full by CLT, and (2) CLT pays the honest cloud $w$ plus a bonus $d - ch$ and refunds its deposit $d$. CLT pays the fee $ch$ to TTP.
(11) If after $T_3 > T_2$, the client has neither paid nor raised dispute, then for any cloud $C_i$ who delivered a result before $T_2$, CLT must pay $C_i$ the agreed amount $w$ and refund its deposit. Any deposit left after that will be transferred to CLT.

In the contract there are various deadlines ($T_1 < T_2 < T_3$). The deadlines are used to enforce timeliness and also to avoid locking away funds if some parties refuse to move forward. The latter is particularly important in smart contracts as the balance in a contract is controlled by a program. Without explicit deadlines and code specifying what to do after the deadlines, the fund can be locked forever by the contract. Note that we assume the client is honest, therefore Clause 11 will never be invoked in this case. The clause is included in the contract to assure the clouds that their funds will not be locked.

Clause 8 says that the client is empowered to settle the contract only when there is an obvious fault, i.e. none of the clouds delivers the result, or when he is satisfied with results. In all other situations, e.g. when only one result is received or the results do not match, the contract must be settled by the TTP. Clauses 9 and 10 deal with the cases in which the TTP is involved. The TTP declares who cheated and then the penalty/reward is dictated by the TTP's judgement. If the client is honest, dispute is only raised when something went wrong and the cost for dispute resolution is covered by the deposit(s) of the cheating cloud(s).

## 5.2 The Game and Analysis

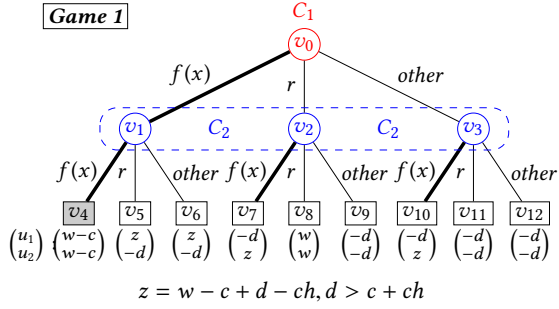

Figure 2: The game induced by the Prisoner's contract. Bold edges indicate the actions that parties will play in the unique sequential equilibrium. The reachable terminal node of the game is in grey.

The game induced by the prisoner contract is shown in Figure 2. In the game, the players are the two clouds, i.e. $\mathcal{N} = \{C_1, C_2\}$. Although the contract also involves the client and the TTP, they can be eliminated from the game because they are honest and have only one deterministic strategy. The clouds can communicate with each other. They can discuss about collusion and work out a value $r \neq f(x)$ that they would both send to cheat the client. In the game, the action set is $\mathcal{A} = \{f(x), r, other\}$. The first two means the party sends $f(x)$ or $r$ before the deadline, the last captures any other actions the party may do. The game has two information sets: $\mathcal{I}_1 = \{v_0\}$ belongs to $C_1$, and $\mathcal{I}_2 = \{v_1, v_2, v_3\}$ belongs to $C_2$. $\mathcal{H}, \mathcal{Z}, \chi, \rho, \sigma$ are captured by the tree structure. We use $u_1$ and $u_2$ to denote $C_1$ and $C_2$'s utility functions respectively. The payoffs (utility) of the parties are listed below the terminal nodes. Table 1 shows how the payoffs are calculated. The table shows which contract clauses are applicable at each terminal node, the payoff for each party prescribed by the contract clauses, the cost of computation and the total amount gained or lost by each party at each terminal node.

Next we analyze the game and show that if the deposit is large enough, more precisely if $d > c + ch$, both parties will always send $f(x)$ and the game will alway ends at $v_4$. We prove by showing that the game has a unique sequential equilibrium in which both parties will play $f(x)$ with a probability 1. Thus, the only reachable outcome is $v_4$. The intuition behind the equilibrium is that for each party, playing $f(x)$ always leads to the highest payoff for itself. Indeed if we look at $C_2$'s decision points $v_1$: $f(x)$ leads to $v_4$ while $r$ leads to $v_5$ and $other$ leads to $v_6$. $C_2$'s payoff is $w - c$ if the game ends at $v_4$ and $-d$ if the game ends at $v_5$ or $v_6$. Since $w - c$ is positive, it is always better than $-d$. Similarly, at decision point $v_2$, $f(x)$ leads to $v_7$. If $d > c + ch$, then $v_7$ has a higher payoff for $C_2$ ($z$) than $v_8$ ($w$) and $v_9$ ($-d$); at decision point $v_3$, $f(x)$ leads to $v_{10}$ that has a higher payoff for $C_2$ ($z$) than $v_{11}$ ($-d$) and $v_{12}$ ($-d$). Therefore $C_2$ will always play $f(x)$ no matter what is $C_1$'s action. Knowing that, $C_1$ knows that the only reachable outcomes are $v_4$, $v_7$ and $v_{10}$

| Outcome | Party | Clause | Payoff in Contract | Cost | Total |
|---|---|---|---|---|---|
| $v_4$ | $C_1$ | 8b | $w$ | $c$ | $w - c$ |
| | $C_2$ | | $w$ | $c$ | $w - c$ |
| $v_5, v_6$ | $C_1$ | 9, 10c | $w + d - ch$ | $c$ | $w - c + d - ch$ |
| | $C_2$ | | $-d$ | 0 | $-d$ |
| $v_7, v_{10}$ | $C_1$ | 9, 10c | $-d$ | 0 | $-d$ |
| | $C_2$ | | $w + d - ch$ | $c$ | $w - c + d - ch$ |
| $v_8$ | $C_1$ | 8b | $w$ | 0 | $w$ |
| | $C_2$ | | $w$ | 0 | $w$ |
| $v_9, v_{11}$ | $C_1$ | 9, 10b | $-d$ | 0 | $-d$ |
| | $C_2$ | | $-d$ | 0 | $-d$ |
| $v_{12}$ | $C_1$ | 8a or | $-d$ | 0 | $-d$ |
| | $C_2$ | (9, 10b) | $-d$ | 0 | $-d$ |

Table 1: Payoff analysis of Game 1

because $C_2$ will never play $r$ or $other$. The payoff at $v_4$ for $C_1$ is $w - c$ that is greater than the payoffs of $v_7$ ($-d$) and $v_{10}$ ($-d$). Thus $C_1$ will choose $f(x)$ in order to reach $v_4$ and get the best payoff. Formally, we have the following:

LEMMA 5.1. *If $d > c + ch$, then Game 1 in Figure 2 has a unique sequential equilibrium $((s_1, s_2), (\beta_1, \beta_2))$ where*

$$\begin{cases} s_1 = ([1(f(x)), 0(r), 0(other)]) \\ s_2 = ([1(f(x)), 0(r), 0(other)]) \\ \beta_1 = ([1(v_0)]) \\ \beta_2 = ([1(v_1), 0(v_2), 0(v_3)]) \end{cases}$$

THEOREM 5.2. *If $d > c + ch$ and $C_1, C_2$ are rational, Game 1 in Figure 2 will always terminate at $v_4$.*

Lemma 5.1 states that the best move for both $C_1$ and $C_2$ is to always send $f(x)$ in time (with a probability 1). Informally, the beliefs can be reasoned as following: for $C_1$, since $\mathcal{I}_1$ has only one node, $C_1$ knows that it is always at $v_0$ when reaching $\mathcal{I}_1$ (i.e. $\beta_1 = ([1(v_0)])$); for $C_2$, knowing that $C_1$'s strategy is to always send $f(x)$, it believes that it always reaches $v_1$ and not the other two nodes in $\mathcal{I}_2$ (i.e. $\beta_2 = ([1(v_1), 0(v_2), 0(v_3)])$). Given Lemma 5.1, Theorem 5.2 can be proved easily: if both parties always send $f(x)$ with a probability 1, the game always ends at $v_4$. The proofs of the Lemma and the Theorem can be found in the Appendix (Section B.1).

## 6 THE COLLUDER'S CONTRACT

The Prisoner's contract works by creating a Prisoner's dilemma between the two clouds. However, it is not strong enough because the dilemma can be solved if the clouds can make credible and enforceable promises. In this section we will show how the two clouds can use another smart contract to counter the Prisoner's contract.

### 6.1 The Contract

In the real world, despite high fines imposed by legal systems, collusion coalitions can still be formed after having an covert agreement among the colluders to redistribute profit and to punish those who deviate from collusion. In the following, we will show the Colluder's contract that captures and enforces such a collusion agreement between the two clouds. Essentially, the Colluder's contract imposes additional rules that will affect the parties' payoffs with the aim to make collusion the most profitable strategy for all colluding parties. In the contract, the cloud who initiates the collusion pays the other cloud a bribe of amount $b$ to incentivize collusion. Also, both clouds pay a deposit of amount $t$ when signing the contract and the party who deviates from collusion will be punished by losing the deposit. The contract is presented below:

(1) The contract should be signed by two clouds $C_1$ and $C_2$. We call the cloud who initiates the collusion the ringleader (LDR). The ringleader can be either $C_1$ or $C_2$. We call the other cloud the follower (FLR).
(2) LDR and FLR agree to deliver a value $r \neq f(x)$ as the computation result in **CTP**, which is a Prisoner's Contract signed by LDR and FLR and a client CLT to compute $f()$ on input $x$.
(3) As a condition, LDR must pay $t + b$ and FLR must pay $t$ when they sign the Colluder's contract. The amount will be paid into and held by the smart contract.
(4) LDR and FLR must pay the amounts stated above before $T_4 <$ **CTP**.$T_2$, where **CTP**.$T_2$ is the result delivery deadline specified in **CTP**. If anyone fails to do so, the contract will terminate and any deposits paid will be refunded.
(5) Once **CTP** has concluded, the following will be done to the balance held by the contract:
    (a) (Both follow) If both LDR and FLR output $r$ in **CTP**, then $t$ is paid to LDR and $t + b$ is paid to FLR;
    (b) (FLR deviates) Else if LDR outputs $r$ in **CTP** and FLR's output in **CTP** is not $r$, then $2 \cdot t + b$ is paid to LDR and FLR gets nothing;
    (c) (LDR deviates) Else if LDR's output is not $r$ in **CTP** and FLR outputs $r$ in **CTP**, then $2 \cdot t + b$ is paid to FLR and LDR gets nothing;
    (d) (Both deviate) Else $t + b$ is paid to LDR and $t$ is paid to FLR.

The contract must be signed before **CTP**.$T_2$ because otherwise it would be too late. The clouds needs to deliver the results in **CTP** (Prisoner's contract) before **CTP**.$T_2$. The collusion agreement must be signed before this time so that the clouds know for sure that the collusion is secured and can deliver $r$ without any risk. In clause 5d, when both clouds deviate from collusion, none of them is punished. Of course, another choice is to punish both in this case. The analysis of this variant is similar and the equilibrium remains the same.

## 6.2 The Game and Analysis

The game induced by the Prisoner's contract and the Colluder's contract is shown in Figure 3. Note that LDR has the choice of not to initiate the collusion coalition, and FLR has the choice of not to collude with LDR. In this two cases, they will not sign the Colluder's contract, and end up playing Game 1 (Figure 2) because the only contract in effect is the Prisoner's contract. We will not show the analysis of these two branches here, as it is exactly the same as we have shown in Section 5.2 (subject to relabelling of nodes). Because only one terminal node is reachable in Game 1, we can replace each branch with a single terminal node, and its payoff is the payoff of the only reachable terminal node in Game 1. Otherwise the payoffs are decided jointly by the Prisoner's contract and the Colluder's contract. The game has four information sets. They are $I_{1,1} = \{v_0\}$ and $I_{1,2} = \{v_2\}$ (belong to LDR), and $I_{2,1} = \{v_1\}$ and $I_{2,2} = \{v_3, v_4, v_5\}$ (belong to FLR). We use $u_1$ and $u_2$ to denote LDR's and FLR's utility functions respectively. he analysis of the payoff can be found in the Appendix (Table 4, Section C.1).

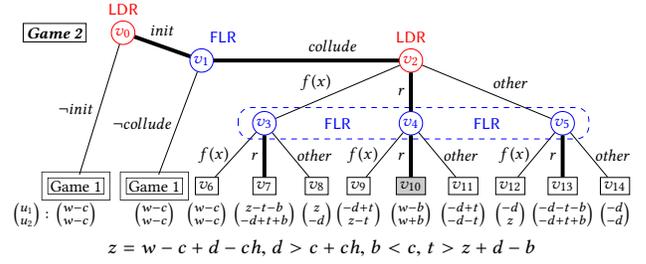

**Figure 3: The game induced by the Prisoner's contract and the Colluder's contract. Bold edges indicate the actions that parties will play in the unique sequential equilibrium. The reachable terminal node of the game is in grey.**

In this contract, LDR pays FLR a bribe for collusion, which needs to satisfy $b < c$, where $c$ is the cost of computing $f(x)$. This is necessary to ensure that LDR has the motivation to initiate the collusion coalition. Note that the collusion is successful if both clouds send $r$. In this case, LDR does not need to compute, but needs to pay a bribe. Its payoff is $w - b$. On the other hand, if there is no collusion and LDR computes honestly, its payoff is $w - c$. Intuitively, LDR would only initiate the collusion coalition if the collusion brings a higher payoff, i.e. when $w - b > w - c$ or equivalently $b < c$. The two clouds also pay a deposit $t$. The amount needs to satisfy $t > z + d - b$, where $z = w - c + d - ch$. This condition is necessary to ensure that (1) the deviating party always gets a payoff no better than what it will get when not deviating, and (2) the party who follows the collusion strategy will always get a higher payoff than not following the strategy. When the conditions are satisfied, we can prove the following Lemma and Theorem:

**LEMMA 6.1.** *If $d > c + ch$, $b < c$ and $t > z + d - b$, then the game in Figure 3 has a unique sequential equilibrium $((s_1, s_2), (\beta_1, \beta_2))$ where $s_1, \beta_1$ are LDR's strategy and beliefs, and $s_2, \beta_2$ are FLR's strategy and beliefs:*

$$\begin{cases} s_1 = ([1(init), 0(\neg init)], [0(f(x)), 1(r), 0(other)]) \\ s_2 = ([1(collude), 0(\neg collude)], [0(f(x)), 1(r), 0(other)]) \\ \beta_1 = ([1(v_0)], [1(v_2)]) \\ \beta_2 = ([1(v_1)], [0(v_3), 1(v_4), 0(v_5)]) \end{cases}$$

**THEOREM 6.2.** *If $d > c + ch$, $b < c$, $t > z + d - b$ and $C_1, C_2$ are rational, Game 2 in Figure 3 will always terminate at $v_{10}$.*

Lemma 6.1 states that the best strategy for LDR is to always initialize the collusion coalition and send $r$ as the result, and the best strategy for FLR is to always collude with LDR and send $r$ as the result. Following the strategies, the game will terminates at $v_{10}$, which gives both clouds the highest payoffs they can get (taking into account the other party's strategy). The proofs of the Lemma and Theorem can be found in the Appendix (Section C.1).

# 7 THE TRAITOR'S CONTRACT

In Section 6 we showed the Colluder's contract that captures and enforces a collusion agreement. The contract enables two clouds to collude and ensures that no one will deviate from collusion. In this section, we show the Traitors' contract, which is designed to address the collusion problem and force the clouds to behave honestly.

## 7.1 The Contract

The main difficulty when designing the Traitor's contract is how to avoid creating a counter/counter-back loop. The client can use a contract to counter the Colluder's contract by providing an additional reward to the honest cloud and change the equilibrium so that collusion is less preferable. However, once the clouds knows what is offered in the contract, they may be able to create a counter contract so that collusion becomes the equilibrium again. This loop can go endlessly.

To get out of the loop, the Traitor's contract works not by countering the Colluder's contract, but by offering the first cloud who reports a collusion to the client the total immunity of the penalty that is imposed by the Prisoner's contract[3]. The aim of the Traitor's contract is not to incentivize the clouds to deviate from the collusion, but to incentivize the clouds to report the collusion. If a Traitor's contract is signed and a collusion is reported, the TTP will step in and decide who cheated. A counter contract is pointless because once the TTP is involved, the payoff of a cloud depends only on whether it cheated but not the other cloud's behavior.

The subtlety of the Traitor's contract is that the immunity granted will allow the reporting cloud to secretly betray the partner while pretending to follow collusion strategy. This is important because without this immunity, a cloud will never report voluntarily: if it reports and follows the collusion strategy, it will lose its deposit in the Prisoner's contract (because TTP will find both clouds are cheating); however if it reports then deviates from the collusion strategy, it will lose its deposit in the Colluder's contract. In either case the reporting cloud is worse off than not reporting. The Traitor's contract promises that the reporting cloud will not be punished by the Prisoner's contract. Then it is safe for the reporting cloud to follow the collusion strategy, and by doing so, the reporting cloud can also get away from the punishment imposed by the Colluder's contract. In consequence, betrayal is risk free. In addition, the Traitor's contract promises a reward to the reporting cloud if the collusion is true. Therefore reporting is preferable to staying in the collusion coalition because it is risk-free and leads to a higher payoff. The Traitor's contract destabilizes collusion by encouraging betrayal. Moreover, the fear of betrayal creates distrust between the clouds. The distrust will eventually deter the formation of the collusion coalition. In addition, the Traitor's contract also punishes misreporting, i.e. a cloud reporting a fabricated case in order to gain benefits. The contract is presented below:

(1) The contract should be signed between a client (CLT) and a cloud who reports collusion. We call this cloud the traitor (TRA). CLT and TRA must have signed **CTP**, a Prisoner's contract.

(2) CLT only signs the Traitor's Contract with the first cloud who reports the collusion. CLT agrees to compensate TRA's loss in **CTP** in suitable cases.

(3) TRA must deliver the computation result of $f(x)$ in this contract, which can be different from the one delivered in **CTP**.

(4) As a condition, CLT must pay a deposit of amount $w + 2 \cdot d - ch$ that equals the maximum amount TRA could lose in **CTP** plus the reward. TRA must pay a deposit of amount $ch$ that equals the fee for dispute resolution. The deposits will be held by the smart contract.

(5) The contract should be fully signed before **CTP**.$T_2$, the deadline for delivering the result in **CTP**. Otherwise the contract terminates and any deposit paid will be refunded.

(6) TRA must deliver a result in this contract before **CTP**.$T_2$.

(7) CLT always raises a dispute instead of invoking Clause 8 in **CTP**.

(8) Once **CTP** is settled by TTP, the following will be done to the deposits held by this contract:
   (a) If in **CTP** none of the clouds cheated (as asserted by TTP), then CLT's deposit $w + 2 \cdot d - ch$ is refunded, and TRA's deposit $ch$ is paid to CLT. Nothing is paid to TRA;
   (b) Else if in **CTP** the other cloud did not cheat and TRA cheated and TRA delivered a correct result in this contract, then $2 \cdot d - ch$ is paid to CLT and $w + ch$ is paid to TRA;
   (c) Else if in **CTP** both clouds cheated and TRA delivered a correct result in this contract, then TRA gets back its deposit $ch$. TRA is also paid $w + 2 \cdot d - ch$. Nothing is paid to CLT;
   (d) Else $w + 2 \cdot d - ch$ is paid to CLT and $ch$ is paid to TRA.

(9) If TRA delivered a result in this contract, and **CTP**.$T_3$ has passed, then all deposits, if any left, go to TRA.

To report collusion, TRA must follow the following procedure:

(i) Wait until the Colluder's contract has been created and signed by the other cloud.
(ii) Before signing the Colluder's contract, report the collusion to the client. Optionally, TRA can submit evidence of collusion e.g. the address of the Colluder's contract and the value $r$ that to be output in the event of collusion.
(iii) Sign the Colluder's contract only after it has signed the Traitor's contract with the client.

CLT only signs the Traitor's contract with the first cloud who reports the collusion. This is because in our case the collusion coalition has only two members. It is too generous to forgive both of them. Once the Traitor's contract is fully signed, CLT always raises a dispute in **CTP**. There are two potential punishments imposed on TRA by the Prisoner's contract and the Colluder's contract. To ensure that TRA's payoff is not worse off in the event of a true collusion, TRA needs to deliver $r$ in **CTP** to get away from the punishment imposed by the Colluder's contract, and then deliver $f(x)$ in the Traitor's contract to get the compensation of the penalty imposed by **CTP** (the Prisoner's contract). It is important that TRA follows the procedure to ensure it signs all three contracts or only **CTP**, otherwise it might have to bear a loss (see Game 3 and Game 4 in the following sections). To dispel TRA's concern of being cheated to "turn in", CLT pays into the contract $w + 2 \cdot d - ch$ to assure TRA that its loss will be compensated and its reward will be given.

---

[3]Technically, the immunity is granted not by exempting the penalty in the Prisoner's contract, but by refunding and compensating the penalty.

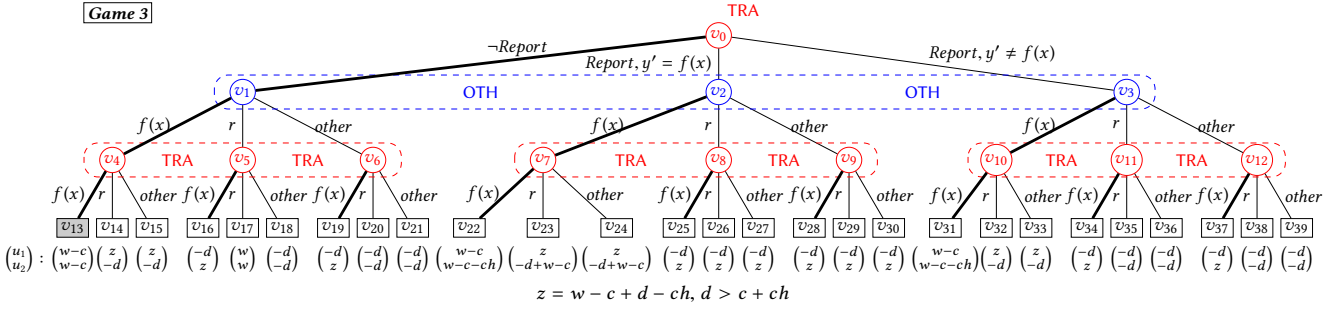

Figure 4: The sub-game induced by the Prisoner's contract and the Traitor's contract. Bold edges indicate the actions that parties will play in the unique sequential equilibrium. The reachable terminal node of the game is in grey.

Before reporting, TRA needs to wait until the other cloud has signed the contract, i.e. fully committed to collusion. Otherwise if TRA reports and the other cloud decides not to sign the Colluder's contract, TRA will be in the situation of (unintentional) misreporting because the other cloud can deliver the correct result in **CTP**. When reporting, TRA can submit evidence of collusion. Note that the evidence submitted by TRA is a "best-effort proof". The purpose of the evidence is not to convince the client about the collusion, but to give the client more information about the collusion. The conclusive evidence of collusion/cheating is TTP's decision and the settlement of Traitor's contract (clause 8) relies only on values in Prisoner's contract and TTP's decision. CLT will sign the Traitor's contract even if the evidence is not strong or verifiable. TRA can falsely report with some fabricated evidence, but as we will show in the next section, a rational cloud will not misreport. This is because when signing the contract, TRA needs to pay $ch$ into the contract and will lose this amount in the event of misreporting.

### 7.2 A Sub-game and Analysis

Before showing the full game, we first show and analyze a sub-game (Game 3, Figure 4). The players in the game include TRA who can be either $C_1$ or $C_2$, and OTH who is the other cloud. We use $u_1$ to denote the utility function of OTH and $u_2$ to denote the utility function of TRA. In Game 3, there is not a fully signed Colluder's contract, either because no one initiates the collusion coalition, or because the collusion attempt is rejected. In the game, TRA can choose not to report at all. If TRA decides not to report (branch to $v_1$), then the only contract in effect is the Prisoner's contract and the branch is exactly the same as the tree of Game 1. On the other hand, TRA can choose to falsely report a case of collusion (misreporting). It also has the choice to later deliver the correct result in the Traitor's contract (branch to $v_2$), or later deliver a wrong result in the Traitor's contract (branch to $v_3$). In both cases, the payoffs of the clouds are affected jointly by the Prisoner's contract and the Traitor's contract. The analysis of the payoffs can be found in the Appendix (Table 5, Section C.2).

The game has five information sets. They are: $\mathcal{I}_1 = \{v_1, v_2, v_3\}$ that belongs to OTH, and $\mathcal{I}_{2,1} = \{v_0\}$, $\mathcal{I}_{2,2} = \{v_4, v_5, v_6\}$, $\mathcal{I}_{2,3} = \{v_7, v_8, v_9\}$ and $\mathcal{I}_{2,4} = \{v_{10}, v_{11}, v_{12}\}$ that belongs to TRA. We have the following Lemma and Theorem:

LEMMA 7.1. *If $d > c + ch$, then Game 3 in Figure 4 has a unique sequential equilibrium $((s_1, s_2), (\beta_1, \beta_2))$ where $s_1, \beta_1$ are OTH's strategy and beliefs, and $s_2, \beta_2$ are TRA's strategy and beliefs:*

$$
\begin{cases}
s_1 = & ([1(f(x)), 0(r), 0(other)]) \\
s_2 = & ([1(\neg report), 0(report, y' = f(x)), 0(report, y' \neq f(x))], \\
& [1(f(x)), 0(r), 0(other)], [1(f(x)), 0(r), 0(other)], \\
& [1(f(x)), 0(r), 0(other)]) \\
\beta_1 = & ([1(v_1), 0(v_2), 0(v_3)]) \\
\beta_2 = & ([1(v_0)], [1(v_4), 0(v_5), 0(v_6)], [1(v_7), 0(v_8), 0(v_9)], \\
& [1(v_{10}), 0(v_{11}), 0(v_{12})])
\end{cases}
$$

THEOREM 7.2. *If $d > c + ch$ and TRA and OTH are rational, then Game 3 in Figure 4 will always terminate at $v_{13}$.*

Lemma 7.1 states that if there is not an effective Colluder's contract, then the best strategy for the clouds is to not report a false collusion case, and to send the correct computation result in the Prisoner's contract. Intuitively, the misreporting cloud will be punished by losing $ch$ and will only end up with a higher payoff if the other cloud happens to cheat. However, without an effective Colluder's contract, the other cloud would unlikely to cheat spontaneously. If the other cloud behaves honestly, then misreporting will lead to a lower payoff than not reporting. Therefore none of the clouds will misreport, and they will send the correct result to get the highest possible payoffs (at $v_{13}$). The proofs can be found in the Appendix (Section C.2).

### 7.3 The Full Game and Analysis

Now we show the full game induced by the three contracts and its analysis. The game is shown in Figure 5. Note that by definition, LDR is the party who initiates the collusion coalition by signing the colluder's contract first, therefore in the game it always moves first. In the game, if LDR decides not to initiate the collusion coalition, or if it initiates but FLR rejects to join, then the two clouds will end up playing Game 3 because there is not a fully signed Colluder's contract. If FLR agrees to collude with LDR, they will enter a different branch. The payoffs in this branch are quite different from those in the Game 3, due to the fact that the Colluder's contract is fully signed and effective. In this branch, it is always FLR who plays the role of traitor, i.e. FLR will be the one that signs the Traitor's contract with the client. If LDR signs the Traitor's contract with the client, then following the report procedure, FLR will not sign the Colluder's contract and the game will go to the ¬*collude* branch. The payoff analysis can be found in the Appendix (Table 6, Section C.3). In the game, there are seven information sets: $I_{1,1} = \{v_0\}$ and $I_{1,2} = \{v_3, v_4, v_5\}$ belong to LDR, $I_{2,1} = \{v_1\}, I_{2,2} = \{v_2\}, I_{2,3} = \{v_6, v_7, v_8\}, I_{2,4} = \{v_9, v_{10}, v_{11}\}$ and $I_{2,5} = \{v_{12}, v_{13}, v_{14}\}$ belong to FLR.

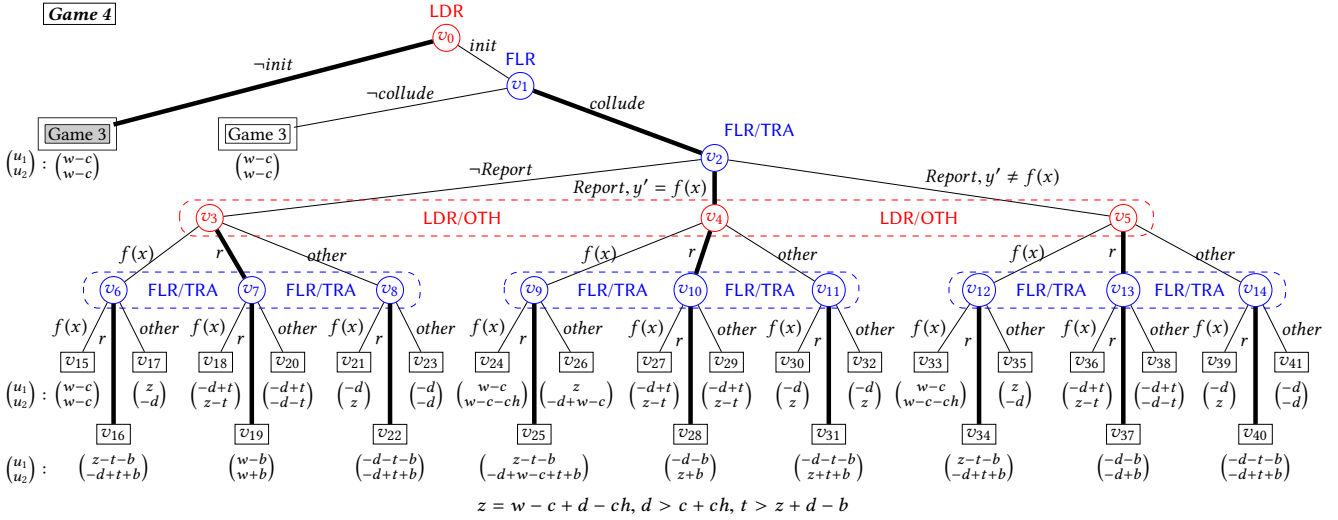

Figure 5: The game induced by the Prisoner's contract, the Colluder's contract and the Traitor's contract. Bold edges indicate the actions that parties will play in the unique sequential equilibrium. The reachable terminal node of the game is in grey.

LEMMA 7.3. *If $d > c + ch, b < c$ and $t > z + d - b$, then Game 4 in Figure 5 has a unique sequential equilibrium $((s_1, s_2), (\beta_1, \beta_2))$ where $s_1, \beta_1$ are LDR's strategy and beliefs, and $s_2, \beta_2$ are FLR's strategy and beliefs:*

$$\begin{cases} s_1 = & ([1(\neg init), 0(init)], [0(f(x)), 1(r), 0(other)]) \\ s_2 = & ([0(\neg collude), 1(collude)], \\ & [0(\neg report), 1(report, y' = f(x)), 0(report, y' \neq f(x))], \\ & [0(f(x)), 1(r), 0(other)], [0(f(x)), 1(r), 0(other)], \\ & [0(f(x)), 1(r), 0(other)]) \\ \beta_1 = & ([1(v_0)], [0(v_3), 1(v_4), 0(v_5)]) \\ \beta_2 = & ([0(v_6), 1(v_7), 0(v_8)], [0(v_9), 1(v_{10}), 0(v_{11})], \\ & [0(v_{12}), 1(v_{13}), 0(v_{14})]) \end{cases}$$

THEOREM 7.4. *If $d > c + ch, b < c$ and $t > z + d - b$ and LDR and FLR are rational, then Game 4 in Figure 5 will always terminate at $v_{13}$ in Game 3 (Figure 4).*

Lemma 7.3 states that in Game 4, LDR (who can be any one of the two clouds) will always choose not to initiate the collusion coalition. The reason that LDR will not attempt to collude is because FLR's best strategy is to pretend to collude and then report the collusion to the client. No matter what LDR does, the payoff it can get from this branch is always less than not to collude. Thus LDR would rather stay away from the collusion. Since no one want to initiate the collusion coalition, there will be no Colluder's contract. Then the two clouds will end up playing Game 3, and the analysis in Section 7.2 shows that they will eventually behave honestly in the sub-game. The proofs can be found in the Appendix (Section C.3).

## 8 IMPLEMENTATION

We implemented the contracts in Solidity 0.4.4 [44] and tested them on the Ethereum network with Geth [18]. We used the Crypto-Con [36], a smart contract that implements elliptic curve cryptography (ECC), for implementing cryptographic operations on blockchain. The contracts are loosely coupled with the actual computation tasks as an external service. The actual computation tasks can be treated as blackboxes and the contracts do not need to know their internal details. The contracts will be called before/during/after executing the tasks, with e.g. the input and output of the tasks. The source code of our contracts can be found at (https://github.com/mjod89/SmartContracts). The pseudocode of the smart contracts and the protocols can be found in Appendix D. We ran the experiments on a MacBook Pro with a 2.8 GHz intel i5 CPU and 8 GB RAM.

### 8.1 Cryptographic Primitives

To implement the contracts on a public blockchain (e.g. Ethereum), we will need to resolve the following challenges:

- Privacy: Since the blockchain is publicly visible to everyone and data on the blockchain is immutable, the biggest concern would be the privacy of input/output of the computation, which need to be specified in the contracts. The client might want to keep them confidential to the public while using the contract.
- Verifiability: While privacy and confidentiality are important, it is also essential that the equality/inequality of the computation results can be verified by the peers in the network because the execution of the contracts is conditioned on those relations.
- Efficiency: The blockchains have a limited space for storing data, and the peers in the network need to verify all transactions. Therefore size and complexity of the transaction are limited.

To address the issues, we use a suitable collision resistant hash function and two other cryptographic primitives: commitments and Non-interactive Zero Knowledge Proofs (NIZK). Informally, a commitment scheme is a two-phase protocol. In the commitment phase, a committer commits to a value $m$ by choosing a secret $s$ to generate a commitment $Com_s(m)$. The commitment should be *hiding*, i.e. it is infeasible to know $m$ given only $Com_s(m)$ but not $s$; the commitment should also be *binding*, i.e. it is infeasible to find $m' \neq m$ and $s' \neq s$ such that $Com_{s'}(m') = Com_s(m)$. In our implementation, we use the well-known Pedersen Commitment Scheme [39]. NIZK allows a prover to non-interactively convince a verifier about a statement without leaking information. We are interested in proving the equality and inequality of values concealed in commitments. More precisely, given two commitments $Com_{s_1}(m_1), Com_{s_2}(m_2)$ and the pairs $(s_1, m_1), (s_2, m_2)$, a prover can generate a proof $\sigma_=$ if $m_1 = m_2$ or $\sigma_{\neq}$ if $m_1 \neq m_2$. Given the commitments and proof, a verifier can run a verification algorithm $V(Com_{s_1}(m_2), Com_{s_2}(m_2), \sigma_=)$ or $V(Com_{s_1}(m_1), Com_{s_2}(m_2), \sigma_{\neq})$ that output 1 only if the relation to be proved holds (expect for a negligible probability). The NIZKs we use are obtained by applying the Fiat-Shamir heuristic on Sigma protocols in [9, 10]. More details can be found in Appendix E.

Instead of using the plaintext input/output, the implementation of the contracts needs to handle cryptographic values and the parties need to run some protocols. The pseudocode of the smart contracts and the protocols can be found in Appendix D. As an example we briefly describe what the parties do when using the Prisoner's contract ($H$ is the collision resistant hash function):

- **Input:** The client chooses the function $f$ (in fact a description or binary code of the function) and input $x$, then computes $m_1 = H(f), m_2 = H(x)$ and two commitments $Com_{s_1}(m_1), Com_{s_2}(m_2)$. The client sends the commitments as part of the contract to the blockchain, and $(f, x, s_1, s_2)$ to the clouds. The clouds verify the commitments on the blockchain is correct then sign the contract.
- **Output:** When delivering the computation result $y_i$, the cloud $C_i$ computes $m_{y_i} = H(y_i)$ and the commitment $Com_{s_{y_i}}(m_{y_i})$. The commitment is sent to the blockchain as part of the transaction and $(y_i, s_{y_i})$ are sent to the client through a private channel.
- **Client proof:** If $y_1, y_2$ received by the client are equal and also the commitments appeared on the blockchain are correct, the client creates a NIZK $\sigma_=$. The Ethereum peers can run the verification algorithm $V(Com_{s_{y_1}}(m_{y_1}), Com_{s_{y_2}}(m_{y_2}), \sigma_=)$ and be convinced if the algorithm outputs 1.
- **TTP proof:** If the client raises a dispute, it sends $(f, x, s_1, s_2)$ and all $(y_i, s_{y_i})$ it received to the TTP. The TTP verifies the commitments on the blockchain are correct, then recomputes $y_t = f(x)$ and computes a commitment $Com_{s_t}(y_t)$. It then compares $y_i$ with $y_t$ to decide who cheated. It then computes an NIZK for each $C_i$ (NULL if $C_i$ didn't deliver a result). If $y_i = y_t$, then the NIZK is $\sigma_=$, otherwise the NIZK is $\sigma_{\neq}$. The TTP sends $Com_{s_t}(y_t)$ and the two NIZK to the blockchain. The peers knows $C_i$ is honest if $V(Com_{s_{y_i}}(m_{y_i}), Com_{s_{y_t}}(m_{y_t}), \sigma_=) = 1$, or $C_i$ is not honest if $C_i$ did not deliver or $V(Com_{s_{y_i}}(m_{y_i}), Com_{s_{y_t}}(m_{y_t}), \sigma_{\neq}) = 1$.

The collision resistance property of the hash function and the binding property of the commitment scheme enable us to replace the actual input/output values that should be put on the blockchain with their commitments. By storing only commitments on the blockchain, we hide information about the input/output. NIZK

| Contract | Functions | Cost in Gas | Cost in $ |
|---|---|---|---|
| Prisoner's | Init | 2,298,950 | 0.4015 |
| | Create | 206,972 | 0.0361 |
| | Bid | 74,899 | 0.0131 |
| | Deliver | 94,373 | 0.0164 |
| | Pay | 821,244 | 0.1434 |
| | Dispute | 2,126,950 | 0.3714 |
| Colluder's | Init | 1,971,270 | 0.3443 |
| | Create | 281,852 | 0.0492 |
| | Join | 58,587 | 0.0102 |
| | Enforce | 103,156 | 0.0180 |
| Traitor's | Init | 2,018,459 | 0.3525 |
| | Create | 161,155 | 0.0281 |
| | Join | 66,802 | 0.0117 |
| | Deliver | 82,846 | 0.0145 |
| | Check | 719,051 | 0.1256 |

Table 2: Cost of using the smart contracts on the official Ethereum network. The transactions are viewable on the blockchain (addresses can be found in Appendix F)

allows the peers to verify equality/inequality of values in the commitments without knowing the actual values. Therefore we solve the privacy problem. The schemes we use are efficient and the overhead is really small (see next section).

## 8.2 Overhead and Cost

**Overhead** The additional overhead incurred by cryptography is small. We implement the commitment and NIZK schemes in ECC. In each contract, each party need to generate at most 2 commitments. Also in each contract at most 2 NIZKs need to be generated and verified. The most costly cryptographic operation is the point multiplication (MUL) operation. Generating a commitment needs 2 MUL. Generating and verifying a equality NIZK each needs 2 MUL as well. Generating an inequality NIZK needs 4 MUL and verifying needs 3 MUL. The commitments and NIZKs are generated locally by the parties. On the blockchain, the peers only need to verify the NIZKs. The commitments and NIZKs are small in size. When using 256-bit ECC, a commitment is only 512 bits, an equality NIZK is 768 bits and an inequality NIZK is 1536 bits. The size can be further reduced if point compression is used.

**Financial Cost** In Table 2, we show the cost of setting up and executing the contracts on the offical Ethereum network. The cost is in the amount of gas consumed by each function, and the converted monetary value in US dollar. The gas price was $2 \times 10^{-9}$ ether (2 Gwei) in all transactions and the exchange rate was 1 ether = \$87.32.

As we can see, the financial cost for using the smart contracts on the Ethereum network is low. The cost is roughly related to the computational and storage complexity of the function. For example, in Prisoner's contract, Init (to store a contract on the blockchain) and Dispute (require verification of NIZKs) cost more than other functions. For the Prisoner's contract, the total cost (for the client and the two clouds) is about 3.8 million gas (\$0.65) if there is no dispute, or about 5 million gas (\$0.88) with dispute resolution. For the Colluder's contract, the total cost is about 2.4 million gas (\$0.42). And for the Traitor's contract, the total cost is about 3 million gas (\$0.53). The cost can be further reduced if the contracts are reused (see Appendix F.2). Note that Ethereum will have native support for ECC [41], which means we can expect a much lower cost for calling functions that involves ECC operations (e.g. Dispute).

## 9 RELATED WORK

**Verifiable computation** There has been much work on verifiable computation based on cryptography, e.g. a good survey can be found in [51] and see [1, 15, 50] for some more recent work. Although providing high-assurance execution, cryptography-based solutions are computationally too costly. The overhead for verification can be made small. However the overhead for pre-computation and for the prover to compute the proof is orders of magnitude higher than the actual cost of the computation being verified. Replication is a long-established technique for building dependable systems, see e.g. [2, 8, 11–13, 26, 32, 43, 47, 49]. To verify the computation, the task is run by $n$ servers and as long as $t$ servers are honest, the correctness of the result can be guaranteed by a consensus protocol. The traditional solution for collusion is to enlarge $n$. The assumption is that collusion will become more difficult or even impossible when $n$ increases. However this is not an option when we have to limit $n$ to 2. In [11], a protocol is designed to allow a client to use a minimum of 2 servers to achieve verifiability. However, the protocol assumes at least one server is honest, thus it precludes the possibility of collusion in the 2 servers case. The protocol also incurs an overhead that is 10 - 20 times higher than the plain execution.

**Game theory and verifiable computation** There has been work on applying game theory in replication based verifiable computation. In [6], the authors considered the 2 servers case and proposed a scheme that induces a game similar to the Prisoner's contract game by punishing the cheating cloud and giving a bounty to the honest cloud when results returned by the clouds do not match. However, in this scheme, the client has to bear the cost of re-computation (to find who cheated) and also the bounty. The penalty paid by the cheating cloud may not be large enough to cover the additional cost to the client. Also as with the Prisoner's contract, the scheme is subject to the collusion attack. To counter collusion, it needs to use multiple servers and assume some of them are honest. The multi-server case is further studied in [31] with an extended scheme. In [38] and [40], the authors proposed similar schemes in which the client gives the task to one cloud, and with a certain probability, also selects another cloud to re-compute the task. To incentivize the clouds to stay honest, contracts were designed to punish the clouds when the results do not match. The schemes are based on a strong assumption that the two clouds cannot communicate, let alone collude, with each other, thus are weaker than the Prisoner's contract. In [25], the authors considered the case in which the clouds can collude (but cannot make creditable and enforceable promises). They proposes contracts that punish both clouds or reward both clouds when the task is not audited and the results are different. The contracts can incentivize honesty. However, if the two clouds can make creditable and enforceable promises (e.g. using a contract similar to the Colluder's contract), they can make collusion the equilibrium of the game.

**Secure Computation with Cryptocurrencies** There is a line of research that focuses on interweaving cryptocurrencies with multiparty secure computation protocols. Most of the work (e.g. [4, 7, 29, 30]) focuses on incentivizing fairness and (timely) delivery of the results. The essential idea is that each party deposits some cryptocurrencis and parties who withhold results will lose their deposits. In [28], the authors considered a crowd-sourcing environment in which a user publishes a job and anyone can submit an answer and gets a bounty. The idea is to use a cryptography-based verifiable computation scheme so that the solver can submit a proof of correctness along with the answer, which will be checked by miners or a designated verifier. The scheme uses cryptocurrencies to solve mainly the fair payment problem, rather than the verifiability problem.

**Rational Adversaries** It has been recognized that in many cases, traditional models of adversaries in cryptography are either too weak (semi-honest) or too strong (malicious). Recently there is a line of research bridging cryptography and game theory that models adversaries as self-interested rational entities [5, 16, 19–22, 24]. The research shows that by considering a rational adversary, which is arguably more realistic, it is possible to design protocols that are more efficient or can circumvent impossibility results.

**Other Related Work** In [37], the authors proposed a method to prevent the concentration of mining power by utilizing distrust. They designed nonoutsourceable puzzles that allow a malicious worker to steal the reward if the mining task is outsourced. The risk would deter mining coalitions such as mining pools or hosted mining. In [48], the authors proposed an attack against mining pools using smart contracts that reward pool workers who withhold their blocks. In [23], the authors showed that smart contracts can be used for malicious purposes. They showed several criminal smart contracts for e.g. leaking confidential information and various real-world crimes.

## 10 CONCLUSION AND FUTURE WORK

Verifiability is a highly desirable property in cloud computing, cost-efficiency is another one. In this paper, we propose a smart contract based solution aiming to achieve both. In our solution, the client outsources the same computation to two clouds, and uses smart contracts to create games between two rational clouds. The games will restrain the clouds from colluding and cutting corners. Instead, they will stay honest to pursue their highest payoffs. Now without collusion, verifiability can be achieved by simply crosschecking the results returned by the clouds. The main cost is the cost for employing two clouds, other costs are small.

In this work we assume the client is honest. One future direction would be to consider the client as a potential adversary. This would make the interplay among parties more complex and requires significant changes to the contracts. Another future direction would be to consider repeated interactions among the parties. Repeated interactions introduces significant changes to the settings because the incentive can be now influenced by reputation and long-term profitability. Also the current deposit mechanism is not very efficient from the cloud point of view. If the cloud has many clients and simultaneous contracts, the cloud must have a large cash reserve to pay all deposits at the same time. One direction would be to investigate more efficient deposit mechanisms by e.g. pooling contracts or insurance. Currently the contracts are written case-by-case. Ultimately we would like to have standard, verified and composable templates/subroutines, much like standard wording/clauses we use in traditional contracts. We would also like to develop counter-collusion contracts in general for other purposes, e.g. to prevent vote buying in e-voting systems like [36].

## ACKNOWLEDGMENTS

The authors would like to thank Dr Helmuts Azacis for the discussion on the initial idea, Prof. Brian Randell for the comments when we preparing the submission and the anonymous reviewers for their valuable comments and helpful suggestions. Changyu Dong and Yilei Wang are supported by the Engineering and Physical Sciences Research Council under Grant No. EP/M013561/2. Patrick McCorry is supported by the Engineering and Physical Sciences Research Council under Grant No. EP/N028104/1.

## A SAVING ON TCO

The following was calculated using the Amazon AWS Total Cost of Ownership (TCO) Calculator on May 3 2017. We used the default assumptions and following configurations:

- Location: US-east (N. Virginia)
- Servers: non-DB, CPU cores per VM = 4, memory per VM =16 GB, Hypervisor = VMware, Guest OS = linux, VM usage =30%, optimized by CPU, Host = 2 CPU, 8 cores, 96 GB RAM.
- Storage: type = SAN, Max IOPS = 1, backup/month = 30%
- Network: data center bandwidth = 1000 Mbit/s, Peak/Average ratio = 3
- IT Labor: Burdened Annual Salary =$120,000, number of VMs per admin = 50

In Table 3 we show the 3-year Total Cost of Ownership for different sizes of IT infrastructure. In the table, small means a small infrastructure with 10 servers and 10 TB storage capacity (with the above configuration), median means 100 servers and 100 TB storage, and large means 1000 servers and 1,000 TB storage.

| Infrastructure Size | On-premises | Cloud | Saving |
|---|---|---|---|
| small | $429,876 | $132,167 | 69% |
| median | $2,112,717 | $980,999 | 54% |
| large | $18,835,526 | $9,356,390 | 50% |

Table 3: 3-year TCO comparison

## B ANALYSIS OF GAMES

### B.1 Analysis of Game 1

#### Proof of Lemma 5.1

PROOF. First, let the strategy profile in the equilibrium be $(s_1, s_2)$ where

$$\begin{cases} s_1 = ([\phi_1(f(x)), \phi_2(r), \phi_3(other)]) \\ s_2 = ([\psi_1(f(x)), \psi_2(r), \psi_3(other)]) \end{cases}$$

In the above, $\phi_i$ and $\psi_i$ are unknown probabilities. They satisfy $0 \le \phi_1, \phi_2, \phi_3 \le 1$ and $\phi_1 + \phi_2 + \phi_3 = 1$, $0 \le \psi_1, \psi_2, \psi_3 \le 1$ and $\psi_1 + \psi_2 + \psi_3 = 1$. In the belief system, $\beta_1 = ([1(v_0)])$ because $\mathcal{I}_1$ has only one node. The beliefs $\beta_2$ can be derived from Bayes' rule:

$$\begin{cases} \beta_1 = ([1(v_0)]) \\ \beta_2 = ([\phi_1(v_1), \phi_2(v_2), \phi_3(v_3)]) \end{cases}$$

Let us reason backwards to find how the players choose their actions. The level above the terminal nodes are three nodes that forms $C_2$'s information set $\mathcal{I}_2 = \{v_1, v_2, v_3\}$. At this information set, it is $C_2$'s turn to move. As a rational player, $C_2$ tries to maximize its expected payoff at this information set, which is:

$$u_2(s; \mathcal{I}_2, \beta) = \beta_2(v_1)u_2(s; v_1) + \beta_2(v_2)u_2(s; v_2) + \beta_2(v_3)u_2(s; v_3)$$
$$= \phi_1 u_2(s; v_1) + \phi_2 u_2(s; v_2) + \phi_3 u_2(s; v_3)$$

In the above, we have:

$$\begin{cases} u_2(s; v_1) = \psi_1 u_2(v_4) + \psi_2 u_2(v_5) + \psi_3 u_2(v_6) \\ u_2(s; v_2) = \psi_1 u_2(v_7) + \psi_2 u_2(v_8) + \psi_3 u_2(v_9) \\ u_2(s; v_3) = \psi_1 u_2(v_{10}) + \psi_2 u_2(v_{11}) + \psi_3 u_2(v_{12}) \end{cases}$$

We argue that if $\psi_1 = 1, \psi_2 = 0, \psi_3 = 0$, i.e. if $C_2$ plays $f(x)$ with a probability 1, then $C_2$ gets the highest expected payoff. Observe that when $d > c + ch$, the following holds: $u_2(v_4) > u_2(v_5) = u_2(v_6)$, $u_2(v_7) > u_2(v_8) > u_2(v_9)$, and $u_2(v_{10}) > u_2(v_{11}) = u_2(v_{12})$. Thus the above probabilities will maximize the expected payoff at each node, i.e. now $u_2(s; v_1) = u_2(v_4), u_2(s; v_2) = u_2(v_7), u_2(s; v_3) = u_2(v_{10})$ are all at their maximum values. In consequence, $u_2(s; \mathcal{I}_2, \beta)$, the expected payoff for $C_2$ at information set $\mathcal{I}_2$, is also maximized.

Now let us move to the level above. It is $C_1$'s turn to move. Since $C_2$'s strategy is $([1(f(x)), 0(r), 0(other)])$, if $C_1$ plays $f(x)$, the outcome will be $v_4$ because $C_2$'s response will be $f(x)$ for sure. Similarly, if $C_1$ plays $r$ the outcome will be $v_7$, and if $C_1$ plays $other$ the outcome will be $v_{10}$. Then the expected payoff of $C_1$ is:

$$u_1(s; \mathcal{I}_1, \beta) = \beta_1(v_0)u_1(s; v_0)$$
$$= \phi_1 u_1(v_4) + \phi_1 u_1(v_7) + \phi_3 u_1(v_{10})$$

In the game, $u_1(v_4) > u_1(v_7) = u_1(v_{10})$. Thus $\phi_1 = 1, \phi_2 = 0, \phi_3 = 0$ will maximize $C_1$'s expected payoff. We can conclude that $E_p$ is sequentially rational because the strategy profile in $E_p$ allows the party to get the maximum payoff at every information set. $E_p$ is also the only sequentially rational assessment because both parties have a strictly dominant strategy.

Consistent can be proven by using the following sequence $s^k = (s_1^k, s_2^k)$ where

$$\begin{cases} s_1^k = \left( \left[ \frac{k-2}{k}(f(x)), \frac{1}{k}(r), \frac{1}{k}(other) \right] \right) \\ s_2^k = \left( \left[ \frac{k-2}{k}(f(x)), \frac{1}{k}(r), \frac{1}{k}(other) \right] \right) \end{cases}$$

It is clear that $s^k$ is fully mixed, i.e. every pure strategy has a non-zero probability. Because $\lim_{k \to \infty} \frac{k-2}{k} = 1$ and $\lim_{k \to \infty} \frac{1}{k} = 0$, $s^k$ converges to $s$. The induced belief system $\beta^k = (\beta_1^k, \beta_2^k)$ is:

$$\begin{cases} \beta_1^k = ([1(v_0)]) \\ \beta_2^k = \left( \left[ \frac{k-2}{k}(v_1), \frac{1}{k}(v_2), \frac{1}{k}(v_3) \right] \right) \end{cases}$$

which also converges to $\beta$. □

#### Proof of Theorem 5.2

PROOF. In Lemma 5.1 we showed that the game has only one sequential equilibrium. In the equilibrium both parties play $f(x)$ with a probability 1, thus the probability of reaching $v_4$:

$$Pr[v_4 | E_p] = Pr[v_0 | s] \cdot Pr[v_1 | (s, v_0)] \cdot Pr[v_4 | (s, v_1)] = 1 \cdot 1 \cdot 1 = 1$$

□

## C ANALYSIS OF GAMES

### C.1 Analysis of Game 2

#### Analysis of Payoffs

See Table 4.

#### Proof of Lemma 6.1

PROOF. First let the strategy profile be $(s_1, s_2)$ where

$$\begin{cases} s_1 = ([\phi_1(\neg init), \phi_2(init)], [\phi_3(f(x)), \phi_4(r), \phi_5(other)]) \\ s_2 = ([\psi_1(\neg collude), \psi_2(collude)], [\psi_3(f(x)), \psi_4(r), \psi_5(other)]) \end{cases}$$

In the above $\phi_i$ and $\psi_i$ are probabilities that satisfy $\phi_1 + \phi_2 = 1$, $\phi_3 + \phi_4 + \phi_5 = 1, \psi_1 + \psi_2 = 1$ and $\psi_3 + \psi_4 + \psi_5 = 1$. Then the belief system is:

$$\begin{cases} \beta_1 = ([1(v_0)], [1(v_2)]) \\ \beta_2 = ([1(v_1)], [\phi_3(v_3), \phi_4(v_4), \phi_5(v_5)]) \end{cases}$$

| Outcome | Party | Prisoner's Contract | | Colluder's Contract | | Cost | Total |
|---|---|---|---|---|---|---|---|
| | | Clause | Payoff in Contract | Clause | Payoff in Contract | | |
| $v_6$ | LDR | 8b | $w$ | 5d | $0$ | $c$ | $w-c$ |
| | FLR | | $w$ | | $0$ | $c$ | $w-c$ |
| $v_7$ | LDR | 9, 10c | $w+d-ch$ | 5c | $-t-b$ | $c$ | $w-c+d-ch-t-b$ |
| | FLR | | $-d$ | | $t+b$ | $0$ | $-d+t+b$ |
| $v_8$ | LDR | 9, 10c | $w+d-ch$ | 5d | $0$ | $c$ | $w-c+d-ch$ |
| | FLR | | $-d$ | | $0$ | $0$ | $-d$ |
| $v_9$ | LDR | 9, 10c | $-d$ | 5b | $t$ | $0$ | $-d+t$ |
| | FLR | | $w+d-ch$ | | $-t$ | $c$ | $w-c+d-ch-t$ |
| $v_{10}$ | LDR | 8b | $w$ | 5a | $-b$ | $0$ | $w-b$ |
| | FLR | | $w$ | | $b$ | $0$ | $w+b$ |
| $v_{11}$ | LDR | 9, 10b | $-d$ | 5b | $t$ | $0$ | $-d+t$ |
| | FLR | | $-d$ | | $-t$ | $0$ | $-d-t$ |
| $v_{12}$ | LDR | 9, 10c | $-d$ | 5d | $0$ | $0$ | $-d$ |
| | FLR | | $w+d-ch$ | | $0$ | $c$ | $w-c+d-ch$ |
| $v_{13}$ | LDR | 9,10b | $-d$ | 5c | $-t-b$ | $0$ | $-d-t-b$ |
| | FLR | | $-d$ | | $t+b$ | $0$ | $-d+t+b$ |
| $v_{14}$ | LDR | 8a or (9,10b) | $-d$ | 5d | $0$ | $0$ | $-d$ |
| | FLR | | $-d$ | | $0$ | $0$ | $-d$ |

Table 4: Payoff analysis of Game 2

Let us reason backward. At information set $\mathcal{I}_{2,2} = \{v_3, v_4, v_5\}$, it is FLR's turn to decide. As a rational player, FLR wants to maximize its expected payoff:

$$u_2(s; \mathcal{I}_{2,2}, \beta) = \beta_2(v_3)u_2(s; v_3) + \beta_2(v_4)u_2(s; v_4) + \beta_2(v_5)u_2(s; v_5)$$
$$= \phi_3 u_2(s; v_3) + \phi_4 u_2(s; v_4) + \phi_5 u_2(s; v_5)$$

In the above we have:

$$\begin{cases} u_2(s; v_3) = \psi_3 u_2(v_6) + \psi_4 u_2(v_7) + \psi_5 u_2(v_8) \\ u_2(s; v_4) = \psi_3 u_2(v_9) + \psi_4 u_2(v_{10}) + \psi_5 u_2(v_{11}) \\ u_2(s; v_5) = \psi_3 u_2(v_{12}) + \psi_4 u_2(v_{12}) + \psi_5 u_2(v_{13}) \end{cases}$$

If $t > z + d - b$, then $u_2(v_7) = -d + t + b > z$. Also, $z = w - c + d - ch > w$ when $d > c + ch$. Therefore $u_2(v_7) > u_2(v_6) > u_2(v_8)$ where $u_2(v_6) = w - c$ and $u_2(v_8) = -d$. Thus $u_2(s; v_3)$ can be maximized if $\psi_3 = 0, \psi_4 = 1$ and $\psi_5 = 0$. Next, $u_2(v_{10}) = w + b$ is greater than $u_2(v_9) = z - t < -d + b$ and $u_2(v_{11}) = -d - t$. Thus $\psi_3 = 0, \psi_4 = 1$ and $\psi_5 = 0$ also maximize $u_2(s; v_4)$. Also $u_2(v_{13}) = -d + b - b > z$ is greater than $u_2(v_{12}) = z$ and $u_2(v_{13}) = -d$. Thus $\psi_3 = 0, \psi_4 = 1$ and $\psi_5 = 0$ also maximize $u_2(s; v_5)$. Therefore to maximize its expected payoff at $\mathcal{I}_{2,2}$, FLR must always play $[0(f(x)), 1(r), 0(other)]$ as part of its strategy. This is regardless of LDR's strategy.

At the level above, it is $\mathcal{I}_{1,2}$ and LDR's turn to play. The expected payoff here is:

$$u_1(s; \mathcal{I}_{1,2}, \beta) = \beta_1(v_1)u_1(s; v_1)$$
$$= \phi_3(\psi_3 u_1(v_6) + \psi_4 u_1(v_7) + \psi_5 u_1(v_8))$$
$$+ \phi_4(\psi_3 u_1(v_9) + \psi_4 u_1(v_{10}) + \psi_5 u_1(v_{11}))$$
$$+ \phi_5(\psi_3 u_1(v_{12}) + \psi_4 u_1(v_{13}) + \psi_5 u_1(v_{14}))$$

Because $\psi_3 = 0, \psi_4 = 1$ and $\psi_5 = 0$, the above can be simplified as $\phi_3 u_1(v_7) + \phi_4 u_1(v_{10}) + \phi_5 u_1(v_{13})$. When $t > z + d - b$, $u_1(v_7) = z - t - b < -d$ and $u_1(v_{13}) = -d - t - b$ are both negative and thus are smaller than $u_1(v_{10}) = w - b$ which is positive because $w > c > b$. Therefore $\phi_3 = 0, \phi_4 = 1$ and $\phi_5 = 0$ maximize $u_1(s; \mathcal{I}_{1,2}, \beta)$. This means a rational LDR must play $[0(f(x)), 1(r), 0(other)]$ as part of its strategy.

A level up, at $\mathcal{I}_{2,1}$, it is FLR's turn to play. Its expected payoff at this information set is $u_2(s; \mathcal{I}_{2,1}, \beta) = \beta_2(v_1)u_2(s; v_1) = \psi_1(u_2(Game\ 1)) + \psi_2(u_2(v_{10}))$. Now $u_2(v_{10}) = w + b$ is greater than $u_2(Game\ 1) = w - c$. So $\psi_1 = 0$ and $\psi_2 = 1$ maximize the expected payoff. This means FLR must play $[0(\neg collude), 1(collude)]$ as part of its strategy.

The top level, at $\mathcal{I}_{1,1}$, it is LDR's turn to play. Its expected payoff at this information set is $u_1(s; \mathcal{I}_{1,1}, \beta) = \beta_1(v_0)u_1(s; v_0) = \phi_1(u_1(Game\ 1)) + \phi_2(u_1(v_{10}))$. If $b < c$, then $u_1(v_{10}) = w - b$ is greater than $u_1(Game\ 1) = w - c$. Then $\phi_1 = 0$ and $\phi_2 = 1$ maximize the expected payoff. This means LDR must play $[0(\neg init), 1(init)]$ as part of its strategy.

Summing up the assessment:

$$\begin{cases} s_1 = ([1(init), 0(\neg init)], [0(f(x)), 1(r), 0(other)]) \\ s_2 = ([1(collude), 0(\neg collude)], [0(f(x)), 1(r), 0(other)]) \\ \beta_1 = ([1(v_0)], [1(v_2)]) \\ \beta_2 = ([1(v_1)], [0(v_3), 1(v_4), 0(v_5)]) \end{cases}$$

always maximizes a party's expected payoff at a information set that belongs to it, thus is sequentially rational.

To show consistent, we use the following sequence $s^k = (s_1^k, s_2^k)$ where

$$\begin{cases} s_1 = \left( [\frac{k-1}{k}(init), \frac{1}{k}(\neg init)], [\frac{1}{k}(f(x)), \frac{k-2}{k}(r), \frac{1}{k}(other)] \right) \\ s_2 = \left( [\frac{k-1}{k}(collude), \frac{1}{k}(\neg collude)], [\frac{1}{k}(f(x)), \frac{k-2}{k}(r), \frac{1}{k}(other)] \right) \end{cases}$$

It is clear that $s^k$ is fully mixed and converge to $s$. The induced belief system:

$$\begin{cases} \beta_1 = ([1(v_0)], 1[(v_2)]) \\ \beta_2 = \left([1(v_1)], [\frac{1}{k}(v_3), \frac{k-2}{k}(v_4), \frac{1}{k}(v_5)]\right) \end{cases}$$

also converges to $\beta$. Thus the assessment is consistent.

Thus the assessment is a sequential equilibrium for the game, and it is the only one. □

### Proof of Theorem 6.2

PROOF. In Lemma 6.1 we showed that the game has only one sequential equilibrium $E_c$. In the equilibrium, LDR will play *init* with a probability 1, followed by FLR playing *collude* with a probability 1, followed by LDR playing $r$ with a probability 1, and followed by FLR playing $r$ with a probability 1:

$$\begin{aligned} Pr[v_{10}|E_c] &= Pr[v_0|s] \cdot Pr[v1|(s,v_0)] \cdot Pr[v2|(s,v_1)] \\ &\quad \cdot Pr[v4|(s,v_2)] \cdot Pr[v_{10}|(s,v_4)] \\ &= 1 \cdot 1 \cdot 1 \cdot 1 \cdot 1 = 1 \end{aligned}$$

□

## C.2 Analysis of Game 3

### Analysis of Payoffs
See Table 5.

### Proof of Lemma 7.1

PROOF. We compute the sequential equilibrium as the following. First let the strategy profile in the equilibrium be $(s_1, s_2)$ where

$$\begin{cases} s_1 = ([\phi_1(f(x)), \phi_2(r), \phi_3(other)]) \\ s_2 = ([\psi_1(\neg report), \psi_2(report, y' = f(x)), \psi_3(report, y' \neq f(x))], \\ \qquad [\psi_4(f(x)), \psi_5(r), \psi_6(other)], [\psi_7(f(x)), \psi_8(r), \psi_9(other)] \\ \qquad [\psi_{10}(f(x)), \psi_{11}(r), \psi_{12}(other)]) \end{cases}$$

In the above, $\phi_i$ and $\psi_i$ are unknown probabilities. They satisfy $0 \leq \phi_1, \phi_2, \phi_3 \leq 1$ and $\phi_1 + \phi_2 + \phi_3 = 1$, $0 \leq \psi_i \leq 1$ and $\psi_1 + \psi_2 + \psi_3 = 1$, $\psi_4 + \psi_5 + \psi_6 = 1$, $\psi_7 + \psi_8 + \psi_9 = 1$ and $\psi_{10} + \psi_{11} + \psi_{12} = 1$. In the belief system:

$$\begin{cases} \beta_1 = ([\psi_1(v_1), \psi_2(v_2), \psi_3(v_3)]) \\ \beta_2 = ([1(v_0)], [\phi_1(v_4), \phi_2(v_5), \phi_3(v_6)], [\phi_1(v_7), \phi_2(v_8), \phi_3(v_9)] \\ \qquad [\phi_1(v_{10}), \phi_2(v_{11}), \phi_3(v_{12})]) \end{cases}$$

Let us start from information set $\mathcal{I}_{2,2} = \{v_4, v_5, v_6\}$. At this information set, it is TRA's turn to move. As a rational player, TRA tries to maximize its expected payoff at this information set, which is:

$$\begin{aligned} u_2(s; \mathcal{I}_{2,2}, \beta) &= \beta_2(v_4)u_2(s; v_4) + \beta_2(v_5)u_2(s; v_5) + \beta_2(v_6)u_2(s; v_6) \\ &= \phi_1 u_2(s; v_4) + \phi_2 u_2(s; v_5) + \phi_3 u_2(s; v_6) \end{aligned}$$

In the above, we have:

$$\begin{cases} u_2(s; v_4) = \psi_4 u_2(v_{13}) + \psi_5 u_2(v_{14}) + \psi_6 u_2(v_{15}) \\ u_2(s; v_5) = \psi_4 u_2(v_{16}) + \psi_5 u_2(v_{17}) + \psi_6 u_2(v_{18}) \\ u_2(s; v_6) = \psi_4 u_2(v_{19}) + \psi_5 u_2(v_{20}) + \psi_6 u_2(v_{21}) \end{cases}$$

We argue that $\psi_4 = 1, \psi_5 = 0, \psi_6 = 0$ when $d > c + ch$. This is because the following holds when $d > c+ch$: $u_2(v_{13}) > u_2(v_{14})$ and $u_2(v_{13}) > u_2(v_6)$, $u_2(v_{16}) > u_2(v_{17}) > u_2(v_{18})$, and $u_2(v_{19}) > u_2(v_{20})$ and $u_2(v_{19}) > u_2(v_{21})$. Thus $\psi_4 = 1, \psi_5 = 0, \psi_6 = 0$ maximize the expected payoff for TRA at this information set.

At information set $\mathcal{I}_{2,3} = \{v_7, v_8, v_9\}$. At this information set, TRA tries to maximize its expected payoff at this information set, which is:

$$\begin{aligned} u_2(s; \mathcal{I}_{2,3}, \beta) &= \beta_2(v_7)u_2(s; v_7) + \beta_2(v_8)u_2(s; v_8) + \beta_2(v_9)u_2(s; v_9) \\ &= \phi_1 u_2(s; v_7) + \phi_2 u_2(s; v_8) + \phi_3 u_2(s; v_9) \end{aligned}$$

In the above, we have:

$$\begin{cases} u_2(s; v_7) = \psi_7 u_2(v_{22}) + \psi_8 u_2(v_{23}) + \psi_9 u_2(v_{24}) \\ u_2(s; v_8) = \psi_7 u_2(v_{25}) + \psi_8 u_2(v_{26}) + \psi_9 u_2(v_{27}) \\ u_2(s; v_9) = \psi_7 u_2(v_{28}) + \psi_8 u_2(v_{29}) + \psi_9 u_2(v_{30}) \end{cases}$$

We argue that $\psi_7 = 1, \psi_8 = 0, \psi_9 = 0$ when $d > c + ch$ and $\phi_1 > 0$. This is because the following holds when $d > c + ch$: $u_2(v_{22}) > u_2(v_{23})$ and $u_2(v_{22}) > u_2(v_{24})$, $u_2(v_{25}) = u_2(v_{26}) = u_2(v_{27})$, and $u_2(v_{28}) = u_2(v_{29}) = u_2(v_{30})$. We can see that (1) $u_2(s; v_7)$ is maximized if $\psi_7 = 1, \psi_8 = 0, \psi_9 = 0$, and (2) $u_2(s; v_8)$ and $u_2(s; v_9)$ are fixed regardless of the value of $\psi_7, \psi_8, \psi_9$. If $\phi_1 = 0$, then any arbitrary $\psi_7, \psi_8, \psi_9$ such that $\psi_7 + \psi_8 + \psi_9 = 1$ can maximize $u_2(s; \mathcal{I}_{2,3}, \beta)$. If $\phi_1 > 0$, then to maximize $u_2(s; \mathcal{I}_{2,3}, \beta)$, we must have $\psi_7 = 1, \psi_8 = 0, \psi_9 = 0$.

At information set $\mathcal{I}_{2,4} = \{v_{10}, v_{11}, v_{12}\}$. At this information set, TRA tries to maximize its expected payoff at this information set, which is:

$$\begin{aligned} u_2(s; \mathcal{I}_{2,4}, \beta) &= \beta_2(v_{10})u_2(s; v_{10}) + \beta_2(v_{11})u_2(s; v_{11}) \\ &\quad + \beta_2(v_{12})u_2(s; v_{12}) \\ &= \phi_1 u_2(s; v_{10}) + \phi_2 u_2(s; v_{11}) + \phi_3 u_2(s; v_{12}) \end{aligned}$$

In the above, we have:

$$\begin{cases} u_2(s; v_{10}) = \psi_{10} u_2(v_{31}) + \psi_{11} u_2(v_{32}) + \psi_{12} u_2(v_{33}) \\ u_2(s; v_{11}) = \psi_{10} u_2(v_{34}) + \psi_{11} u_2(v_{35}) + \psi_{12} u_2(v_{36}) \\ u_2(s; v_{12}) = \psi_{10} u_2(v_{37}) + \psi_{11} u_2(v_{38}) + \psi_{12} u_2(v_{39}) \end{cases}$$

We argue that $\psi_{10} = 1, \psi_{11} = 0, \psi_{12} = 0$ when $d > c + ch$. This is because the following holds when $d > c + ch$: $u_2(v_{31})$ is greater than both $u_2(v_{32})$ and $u_2(v_{33})$, $u_2(v_{34})$ is greater than both $u_2(v_{35})$ and $u_2(v_{36})$, and $u_2(v_{37})$ is greater than both $u_2(v_{38})$ and $u_2(v_{38})$. Thus $\psi_{10} = 1, \psi_{11} = 0, \psi_{12} = 0$ maximize $u_2(s; v_{10}), u_2(s; v_{11})$ and $u_2(s; v_{12})$, and in turn maximize $u_2(s; \mathcal{I}_{2,4}, \beta)$.

Now let us go one level above to information set $\mathcal{I}_1 = \{v_1, v_2, v_3\}$. It is OTH's turn to move. OTH will try to maximize its expected payoff:

$$\begin{aligned} u_1(s; \mathcal{I}_1, \beta) &= \beta_1(v_1)u_1(s; v_1) + \beta_1(v_2)u_1(s; v_2) + \beta_3(v_3)u_1(s; v_3) \\ &= \psi_1 u_1(s; v_1) + \psi_2 u_1(s; v_2) + \psi_3 u_1(s; v_3) \end{aligned}$$

In the above, we have:

$$\begin{cases} u_1(s; v_1) = \phi_1(\psi_4 u_1(v_{13}) + \psi_5 u_1(v_{14}) + \psi_6 u_1(v_{15})) + \\ \qquad \phi_2(\psi_4 u_1(v_{16}) + \psi_5 u_1(v_{17}) + \psi_6 u_1(v_{18})) + \\ \qquad \phi_3(\psi_4 u_1(v_{19}) + \psi_5 u_1(v_{20}) + \psi_6 u_1(v_{21})) \\ u_1(s; v_2) = \phi_1(\psi_7 u_1(v_{22}) + \psi_8 u_1(v_{23}) + \psi_9 u_1(v_{24})) + \\ \qquad \phi_2(\psi_7 u_1(v_{25}) + \psi_8 u_1(v_{26}) + \psi_9 u_1(v_{27})) + \\ \qquad \phi_3(\psi_7 u_1(v_{28}) + \psi_8 u_1(v_{29}) + \psi_9 u_1(v_{30})) \\ u_1(s; v_3) = \phi_1(\psi_{10} u_1(v_{31}) + \psi_{11} u_1(v_{32}) + \psi_{12} u_1(v_{33})) + \\ \qquad \phi_2(\psi_{10} u_1(v_{34}) + \psi_{11} u_1(v_{35}) + \psi_{12} u_1(v_{36})) + \\ \qquad \phi_3(\psi_{10} u_1(v_{37}) + \psi_{11} u_1(v_{38}) + \psi_{12} u_1(v_{39})) \end{cases}$$

We can reason as the following

| Outcome | Party | Prisoner's Contract | | Traitor's Contract | | Cost | Total |
|---|---|---|---|---|---|---|---|
| | | Clause | Payoff in Contract | Clause | Payoff in Contract | | |
| $v_{13}$ | OTH | 8b | $w$ | N/A | | $c$ | $w-c$ |
| | TRA | | $w$ | | | $c$ | $w-c$ |
| $v_{14}, v_{15}$ | OTH | 9, 10c | $w+d-ch$ | N/A | | $c$ | $w-c+d-ch$ |
| | TRA | | $-d$ | | | 0 | $-d$ |
| $v_{16}, v_{19}$ | OTH | 9, 10c | $-d$ | N/A | | 0 | $-d$ |
| | TRA | | $w+d-ch$ | | | $c$ | $w-c+d-ch$ |
| $v_{17}$ | OTH | 8b | $w$ | N/A | | 0 | $w$ |
| | TRA | | $w$ | | | 0 | $w$ |
| $v_{18}, v_{20}$ | OTH | 9, 10b | $-d$ | N/A | | 0 | $-d$ |
| | TRA | | $-d$ | | | 0 | $-d$ |
| $v_{21}$ | OTH | 8a or (9, 10b) | $-d$ | N/A | | 0 | $-d$ |
| | TRA | | $-d$ | | | 0 | $-d$ |
| $v_{22}, v_{31}$ | OTH | 10a | $w$ | 8a | | 0 | $w-c$ |
| | TRA | | $w$ | | $-ch$ | 0 | $w-c-ch$ |
| $v_{23}, v_{24}$ | OTH | 9, 10c | $w+d-ch$ | 8b | | $c$ | $w-c+d-ch$ |
| | TRA | | $-d$ | | $w$ | $c$ | $-d+w-c$ |
| $v_{25} - v_{29}$ | OTH | 9, 10b | $-d$ | 8c | | 0 | $-d$ |
| | TRA | | $-d$ | | $w+2 \cdot d-ch$ | $c$ | $w-c+d-ch$ |
| $v_{30}$ | OTH | 8a or (9, 10b) | $-d$ | 8c | | 0 | $-d$ |
| | TRA | | $-d$ | | $w+2 \cdot d-ch$ | $c$ | $w-c+d-ch$ |
| $v_{32}, v_{33}$ | OTH | 9, 10c | $w+d-ch$ | 8d | | $c$ | $w-c+d-ch$ |
| | TRA | | $-d$ | | 0 | 0 | $-d$ |
| $v_{34}, v_{37}$ | OTH | 9, 10c | $-d$ | 8d | | 0 | $-d$ |
| | TRA | | $w+d-ch$ | | 0 | $c$ | $w-c+d-ch$ |
| $v_{35}, v_{36}, v_{38}$ | OTH | 9, 10b | $-d$ | 8d | | 0 | $-d$ |
| | TRA | | $-d$ | | 0 | 0 | $-d$ |
| $v_{39}$ | OTH | 8a or (9, 10b) | $-d$ | 8d | | 0 | $-d$ |
| | TRA | | $-d$ | | 0 | 0 | $-d$ |

Table 5: Payoff analysis of Game 3

- Since $\psi_4 = 1, \psi_5 = 0, \psi_6 = 0$, $u_1(s, v_1) = \phi_1 u_1(v_{13}) + \phi_2 u_1(v_{16}) + \phi_3 u_1(v_{19})$. Because $u_1(v_{13})$ is greater than $u_1(v_{16})$ and $u_1(v_{19})$, $\phi_1 = 1, \phi_2 = 0, \phi_3 = 0$ maximize $u_1(s, v_1)$.
- Similarly, since $\psi_{10} = 1, \psi_{11} = 0, \psi_{12} = 0$, $u_1(s, v_3) = \phi_1 u_1(v_{31}) + \phi_2 u_1(v_{34}) + \phi_3 u_1(v_{37})$. When $d > c + ch$ $u_1(v_{31})$ is greater than $u_1(v_{34})$ and $u_1(v_{37})$, then $\phi_1 = 1, \phi_2 = 0, \phi_3 = 0$ also maximize $u_1(s, v_3)$.
- For $u_1(s, v_2)$ we have two cases when $\phi_1 > 0$ and when $\phi_1 = 0$:
  (1) In the case of $\phi_1 > 0$, then $\psi_7 = 1, \psi_8 = 0, \psi_9 = 0$. In this case $\phi_1 = 1, \phi_2 = 0, \phi_3 = 0$ maximize $u_1(s, v_2)$ and OTH's payoff will be $w - c$.
  (2) In the case of $\phi_1 = 0$, then $u_1(s, v_2) = -d$. This payoff is less than $w - c$ in the case above.

Therefore to maximize $u_1(s, v_2)$, OTH must choose $\phi_1 = 1, \phi_2 = 0, \phi_3 = 0$.

Summing up, it is easy to see that to maximize $u_1(s; \mathcal{I}_1, \beta)$, OTH must choose $\phi_1 = 1, \phi_2 = 0, \phi_3 = 0$.

Now going up, at information set $\mathcal{I}_{2,1} = \{v_0\}$, it is TRA's turn to move. Its expected payoff:

$$\begin{aligned} u_2(s; \mathcal{I}_{2,1}, \beta) &= \beta_2(v_0) u_2(s; v_0) \\ &= u_2(s; v_0) \\ &= \psi_1 u_2(v_{13}) + \psi_2 u_2(v_{22}) + \psi_3 u_2(v_{31}) \end{aligned}$$

Because $u_2(v_{13})$ is greater than $u_2(v_{22})$ and $u_2(v_{31})$, TRA must choose $\psi_1 = 1, \psi_2 = 0, \psi_3 = 0$ to maximize its expected payoff.

Consistent can be proven by using the following sequence $s^k = (s_1^k, s_2^k)$ where

$$\begin{cases} s_1^k = & \left( [\frac{k-2}{k}(f(x)), \frac{1}{k}(r), \frac{1}{k}(other)] \right) \\ s_2^k = & ([\frac{k-2}{k}(\neg report), \frac{1}{k}(report, y' = f(x)), \frac{1}{k}(report, y' \neq f(x))], \\ & [\frac{k-2}{k}(f(x)), \frac{1}{k}(r), \frac{1}{k}(other)], [\frac{k-2}{k}(f(x)), \frac{1}{k}(r), \frac{1}{k}(other)], \\ & [\frac{k-2}{k}(f(x)), \frac{1}{k}(r), \frac{1}{k}(other)]) \end{cases}$$

It is clear that $s^k$ is fully mixed, i.e. every pure strategy has a non-zero probability. Because $\lim_{k \to \infty} \frac{k-2}{k} = 1$ and $\lim_{k \to \infty} \frac{1}{k} = 0$,

$s^k$ converges to $s$. The induced belief system $\beta^k = (\beta_1^k, \beta_2^k)$ is:

$$\begin{cases} \beta_1^k = & \left([\frac{k-2}{k}(v_1), \frac{1}{k}(v_2), \frac{1}{k}(v_3)]\right) \\ \beta_2^k = & ([1(v_0)], [\frac{k-2}{k}(v_4), \frac{1}{k}(v_5), \frac{1}{k}(v_6)], [\frac{k-2}{k}(v_7), \frac{1}{k}(v_8), \frac{1}{k}(v_9)] \\ & [\frac{k-2}{k}(v_{10}), \frac{1}{k}(v_{11}), \frac{1}{k}(v_{12})]) \end{cases}$$

which also converges to $\beta$. □

### Proof of Theorem 7.2

PROOF. In Lemma 7.1 we showed that the game has only one sequential equilibrium. In the equilibrium both parties play $f(x)$ with a probability 1, thus the probability of reaching $v_4$:

$$\begin{aligned} Pr[v_{13}|E_c] &= Pr[v_0|s] \cdot Pr[v_1|(s, v_0)] \cdot Pr[v_4|(s, v_1)] \cdot Pr[v_{13}|(s, v_4)] \\ &= 1 \cdot 1 \cdot 1 \cdot 1 = 1 \end{aligned}$$

□

## C.3 Analysis of Game 4

### Analysis of Payoffs

See Table 6.

### Proof of Lemma 7.3

PROOF. First let the strategy profile in the equilibrium be $(s_1, s_2)$ where

$$\begin{cases} s_1 = & ([\phi_1(\neg init), \phi_2(init)], [\phi_3(f(x)), \phi_4(r), \phi_5(other)]) \\ s_2 = & ([\psi_1(\neg collude), \psi_2(collude)], \\ & [\psi_3(\neg report), \psi_4(report, y' = f(x)), \psi_5(report, y' \neq f(x))], \\ & [\psi_6(f(x)), \psi_7(r), \psi_8(other)], [\psi_9(f(x)), \psi_{10}(r), \psi_{11}(other)], \\ & [\psi_{12}(f(x)), \psi_{13}(r), \psi_{14}(other)]) \end{cases}$$

In the above, $\phi_i$ and $\psi_i$ are probabilities. They satisfy $0 \leq \phi_i \leq 1$ and $\phi_1 + \phi_2 = 1$, $\phi_3 + \phi_4 + \phi_5 = 1$, $0 \leq \psi_i \leq 1$ and $\psi_1 + \psi_2 = 1$, $\psi_3 + \psi_4 + \psi_5 = 1$, $\psi_6 + \psi_7 + \psi_8 = 1$, $\psi_9 + \psi_{10} + \psi_{11} = 1$ and $\psi_{12} + \psi_{13} + \psi_{14} = 1$. In the belief system:

$$\begin{cases} \beta_1 = & ([1(v_0)], [\psi_3(v_3), \psi_4(v_4), \psi_5(v_5)]) \\ \beta_2 = & ([1(v_1)], [1(v_2)], [\phi_3(v_6), \phi_4(v_7), \phi_5(v_8)], \\ & [\phi_3(v_9), \phi_4(v_{10}), \phi_5(v_{11})], [\phi_3(v_{12}), \phi_4(v_{13}), \phi_5(v_{14})]) \end{cases}$$

Let us start from information set $\mathcal{I}_{2,3} = \{v_6, v_7, v_8\}$. At this information set, it is FLR's turn to move. As a rational player, FLR tries to maximize its expected payoff at this information set, which is:

$$\begin{aligned} u_2(s; \mathcal{I}_{2,3}, \beta) &= \beta_2(v_6)u_2(s; v_6) + \beta_2(v_7)u_2(s; v_7) + \beta_2(v_8)u_2(s; v_8) \\ &= \phi_3 u_2(s; v_6) + \phi_4 u_2(s; v_7) + \phi_5 u_2(s; v_8) \end{aligned}$$

In the above, we have:

$$\begin{cases} u_2(s; v_6) = \psi_6 u_2(v_{15}) + \psi_7 u_2(v_{16}) + \psi_8 u_2(v_{17}) \\ u_2(s; v_7) = \psi_6 u_2(v_{18}) + \psi_7 u_2(v_{19}) + \psi_8 u_2(v_{20}) \\ u_2(s; v_8) = \psi_6 u_2(v_{21}) + \psi_7 u_2(v_{22}) + \psi_8 u_2(v_{23}) \end{cases}$$

We argue that $\psi_6 = 0, \psi_7 = 1, \psi_8 = 0$. This is because the following holds when $d > c + ch, b < c$ and $t > z + d - b$: $u_2(v_{16}) > u_2(v_{15}) > u_2(v_{17})$, $u_2(v_{19}) > u_2(v_{18}) > u_2(v_{20})$, and $u_2(v_{22}) > u_2(v_{21}) > u_2(v_{23})$. Thus $\psi_6 = 0, \psi_7 = 1, \psi_8 = 0$ maximize the expected payoff for FLR at this information set.

At information set $\mathcal{I}_{2,4} = \{v_9, v_{10}, v_{11}\}$. At this information set, FLR tries to maximize its expected payoff at this information set, which is:

$$\begin{aligned} & u_2(s; \mathcal{I}_{2,4}, \beta) \\ =& \beta_2(v_9)u_2(s; v_9) + \beta_2(v_{10})u_2(s; v_{10}) + \beta_2(v_{11})u_2(s; v_{11}) \\ =& \phi_3 u_2(s; v_9) + \phi_4 u_2(s; v_{10}) + \phi_5 u_2(s; v_{11}) \end{aligned}$$

In the above, we have:

$$\begin{cases} u_2(s; v_9) = \psi_9 u_2(v_{24}) + \psi_{10} u_2(v_{25}) + \psi_{11} u_2(v_{26}) \\ u_2(s; v_{10}) = \psi_9 u_2(v_{27}) + \psi_{10} u_2(v_{28}) + \psi_{11} u_2(v_{29}) \\ u_2(s; v_{11}) = \psi_9 u_2(v_{30}) + \psi_{10} u_2(v_{31}) + \psi_{11} u_2(v_{32}) \end{cases}$$

We argue that $\psi_9 = 0, \psi_8 = 1, \psi_9 = 0$. This is because the following holds when $d > c + ch, b < c$ and $t > z + d - b$: $u_2(v_{25}) > u_2(v_{24}) > u_2(v_{26})$, $u_2(v_{28}) > u_2(v_{27}) = u_2(v_{29})$ and $u_2(v_{31}) > u_2(v_{30}) = u_2(v_{32})$. Thus $\psi_9 = 0, \psi_8 = 1, \psi_9 = 0$ maximize the expected payoff for FLR at this information set.

At information set $\mathcal{I}_{2,5} = \{v_{12}, v_{13}, v_{14}\}$. At this information set, FLR tries to maximize its expected payoff at this information set, which is:

$$\begin{aligned} & u_2(s; \mathcal{I}_{2,5}, \beta) \\ =& \beta_2(v_{12})u_2(s; v_{12}) + \beta_2(v_{13})u_2(s; v_{13}) + \beta_2(v_{14})u_2(s; v_{14}) \\ =& \phi_3 u_2(s; v_{12}) + \phi_4 u_2(s; v_{13}) + \phi_5 u_2(s; v_{14}) \end{aligned}$$

In the above, we have:

$$\begin{cases} u_2(s; v_{12}) = \psi_{12} u_2(v_{33}) + \psi_{13} u_2(v_{34}) + \psi_{14} u_2(v_{35}) \\ u_2(s; v_{13}) = \psi_{12} u_2(v_{36}) + \psi_{13} u_2(v_{37}) + \psi_{14} u_2(v_{38}) \\ u_2(s; v_{14}) = \psi_{12} u_2(v_{39}) + \psi_{13} u_2(v_{40}) + \psi_{14} u_2(v_{41}) \end{cases}$$

We argue that $\psi_{12} = 0, \psi_{13} = 1, \psi_{14} = 0$. This is because the following holds when $d > c + ch, b < c$ and $t > z + d - b$: $u_2(v_{34}) > u_2(v_{33}) > u_2(v_{35})$, $u_2(v_{37}) > u_2(v_{36}) > u_2(v_{38})$, and $u_2(v_{40}) > u_2(v_{39}) > u_2(v_{41})$. Thus $\psi_{12} = 0, \psi_{13} = 1, \psi_{14} = 0$ maximize the expected payoff for FLR at this information set.

Now let us go one level above to information set $\mathcal{I}_{1,2} = \{v_3, v_4, v_5\}$. It is LDR's turn to move. LDR will try to maximize its expected payoff:

$$\begin{aligned} u_1(s; \mathcal{I}_{1,2}, \beta) &= \beta_1(v_3)u_1(s; v_3) + \beta_1(v_4)u_1(s; v_4) + \beta_3(v_5)u_1(s; v_5) \\ &= \psi_3 u_1(s; v_3) + \psi_4 u_1(s; v_4) + \psi_5 u_1(s; v_5) \end{aligned}$$

In the above, we have:

$$\begin{cases} u_1(s; v_3) = & \phi_3(\psi_6 u_1(v_{15}) + \psi_7 u_1(v_{16}) + \psi_8 u_1(v_{17})) + \\ & \phi_4(\psi_6 u_1(v_{18}) + \psi_7 u_1(v_{19}) + \psi_8 u_1(v_{20})) + \\ & \phi_5(\psi_6 u_1(v_{21}) + \psi_7 u_1(v_{22}) + \psi_8 u_1(v_{23})) \\ u_1(s; v_4) = & \phi_3(\psi_9 u_1(v_{24}) + \psi_{10} u_1(v_{25}) + \psi_{11} u_1(v_{26})) + \\ & \phi_4(\psi_9 u_1(v_{27}) + \psi_{10} u_1(v_{28}) + \psi_{11} u_1(v_{29})) + \\ & \phi_5(\psi_9 u_1(v_{30}) + \psi_{10} u_1(v_{31}) + \psi_{11} u_1(v_{32})) \\ u_1(s; v_5) = & \phi_3(\psi_{12} u_1(v_{33}) + \psi_{13} u_1(v_{34}) + \psi_{14} u_1(v_{35})) + \\ & \phi_4(\psi_{12} u_1(v_{36}) + \psi_{13} u_1(v_{37}) + \psi_{14} u_1(v_{38})) + \\ & \phi_5(\psi_{12} u_1(v_{39}) + \psi_{13} u_1(v_{40}) + \psi_{14} u_1(v_{41})) \end{cases}$$

We can reason as the following

- Since $\psi_6 = 0, \psi_7 = 1, \psi_8 = 0$, $u_1(s, v_3) = \phi_3 u_1(v_{16}) + \phi_4 u_1(v_{19}) + \phi_5 u_1(v_{22})$. Because $u_1(v_{19}) > u_1(v_{16}) > u_1(v_{22})$, $\phi_3 = 0, \phi_4 = 1, \phi_5 = 0$ maximize $u_1(s, v_3)$.
- Similarly, since $\psi_9 = 0, \psi_{10} = 1, \psi_{11} = 0$, $u_1(s, v_4) = \phi_3 u_1(v_{25}) + \phi_4 u_1(v_{28}) + \phi_5 u_1(v_{31})$. Because $u_1(v_{28}) > u_1(v_{25}) > u_1(v_{31})$, $\phi_3 = 0, \phi_4 = 1, \phi_5 = 0$ also maximize $u_1(s, v_4)$.
- Similarly, since $\psi_{12} = 0, \psi_{13} = 1, \psi_{14} = 0$, $u_1(s, v_5) = \phi_3 u_1(v_{34}) + \phi_4 u_1(v_{37}) + \phi_5 u_1(v_{40})$. Because $u_1(v_{37}) > u_1(v_{34}) > u_1(v_{40})$, $\phi_3 = 0, \phi_4 = 1, \phi_5 = 0$ also maximize $u_1(s, v_5)$.

Summing up, it is easy to see that to maximize $u_1(s; \mathcal{I}_{1,2}, \beta)$, OTH must choose $\phi_3 = 0, \phi_4 = 1, \phi_5 = 0$.

| Outcome | Party | Prisoner's Contract | | Colluder's Contract | | Traitor's Contract | | Cost | Total |
|---|---|---|---|---|---|---|---|---|---|
| | | Clause | Payoff in Contract | Clause | Payoff in Contract | Clause | Payoff in Contract | | |
| $v_{15}$ | LDR | 8b | $w$ | 5d | $0$ | N/A | | $c$ | $w-c$ |
| | FLR | | $w$ | | $0$ | | | $c$ | $w-c$ |
| $v_{16}$ | LDR | 9, 10c | $w+d-ch$ | 5c | $-t-b$ | N/A | | $c$ | $w-c+d-ch-t-b$ |
| | FLR | | $-d$ | | $t+b$ | | | $0$ | $-d+t+b$ |
| $v_{17}$ | LDR | 9, 10c | $w+d-ch$ | 5d | $0$ | N/A | | $c$ | $w-c+d-ch$ |
| | FLR | | $-d$ | | $0$ | | | $0$ | $-d$ |
| $v_{18}$ | LDR | 9, 10c | $-d$ | 5b | $t$ | N/A | | $0$ | $-d+t$ |
| | FLR | | $w+d-ch$ | | $-t$ | | | $c$ | $w-c+d-ch-t$ |
| $v_{19}$ | LDR | 8b | $w$ | 5a | $-b$ | N/A | | $0$ | $w-b$ |
| | FLR | | $w$ | | $b$ | | | $0$ | $w+b$ |
| $v_{20}$ | LDR | 9, 10b | $-d$ | 5b | $t$ | N/A | | $0$ | $-d+t$ |
| | FLR | | $-d$ | | $-t$ | | | $0$ | $-d-t$ |
| $v_{21}$ | LDR | 9, 10c | $-d$ | 5d | $0$ | N/A | | $0$ | $-d$ |
| | FLR | | $w+d-ch$ | | $-t$ | | | $c$ | $w-c+d-ch$ |
| $v_{22}$ | LDR | 9,10b | $-d$ | 5c | $-t-b$ | N/A | | $0$ | $-d-t-b$ |
| | FLR | | $-d$ | | $t+b$ | | | $0$ | $-d+t+b$ |
| $v_{23}$ | LDR | 8a or 9,10b | $-d$ | 5d | $0$ | N/A | | $0$ | $-d$ |
| | FLR | | $-d$ | | $0$ | | | $0$ | $-d$ |
| $v_{24}, v_{33}$ | LDR | 10a | $w$ | 5d | $0$ | 8a | | $c$ | $w-c$ |
| | FLR | | $w$ | | $0$ | | $-ch$ | $c$ | $w-c-ch$ |
| $v_{25}$ | LDR | 9, 10c | $w+d-ch$ | 5c | $-t-b$ | 8b | | $c$ | $w-c+d-ch-t-b$ |
| | FLR | | $-d$ | | $t+b$ | | $w$ | $c$ | $-d+w-c+t+b$ |
| $v_{26}$ | LDR | 9, 10c | $w+d-ch$ | 5d | $0$ | 8b | | $c$ | $w-c+d-ch$ |
| | FLR | | $-d$ | | $0$ | | $w$ | $c$ | $-d+w-c$ |
| $v_{27}$ | LDR | 9, 10c | $-d$ | 5b | $t$ | 8d | | $0$ | $-d+t$ |
| | FLR | | $w+d-ch$ | | $-t$ | | $0$ | $c$ | $w-c+d-ch-t$ |
| $v_{28}$ | LDR | 9, 10b | $-d$ | 5a | $-b$ | 8c | | $0$ | $-d-b$ |
| | FLR | | $-d$ | | $b$ | | $w+2\cdot d-ch$ | $c$ | $w-c+d-ch+b$ |
| $v_{29}$ | LDR | 9, 10b | $-d$ | 5b | $t$ | 8c | | $0$ | $-d+t$ |
| | FLR | | $-d$ | | $-t$ | | $w+2\cdot d-ch$ | $c$ | $w-c+d-ch-t$ |
| $v_{30}$ | LDR | 9, 10c | $-d$ | 5d | $0$ | 8d | | $0$ | $-d$ |
| | FLR | | $w+d-ch$ | | $0$ | | $0$ | $c$ | $w-c+d-ch$ |
| $v_{31}$ | LDR | 9, 10b | $-d$ | 5c | $-t-b$ | 8c | | $0$ | $-d-t-b$ |
| | FLR | | $-d$ | | $t+b$ | | $w+2\cdot d-ch$ | $c$ | $w-c+d-ch+t+b$ |
| $v_{32}$ | LDR | 8a or 9, 10b | $-d$ | 5d | $0$ | 8c | | $0$ | $-d$ |
| | FLR | | $-d$ | | $0$ | | $w+2\cdot d-ch$ | $c$ | $w-c+d-ch$ |
| $v_{34}$ | LDR | 9, 10c | $w+d-ch$ | 5c | $-t-b$ | 8d | | $c$ | $w-c+d-ch-t-b$ |
| | FLR | | $-d$ | | $t+b$ | | $0$ | $0$ | $-d+t+b$ |
| $v_{35}$ | LDR | 9, 10c | $w+d-ch$ | 5d | $0$ | 8d | | $c$ | $w-c+d-ch$ |
| | FLR | | $-d$ | | $0$ | | $0$ | $0$ | $-d$ |
| $v_{36}$ | LDR | 9, 10c | $-d$ | 5b | $t$ | 8d | | $0$ | $-d+t$ |
| | FLR | | $w+d-ch$ | | $-t$ | | $0$ | $c$ | $w-c+d-ch-t$ |
| $v_{37}$ | LDR | 9, 10b | $-d$ | 5a | $-b$ | 8d | | $0$ | $-d-b$ |
| | FLR | | $-d$ | | $b$ | | $0$ | $0$ | $-d+b$ |
| $v_{38}$ | LDR | 9, 10b | $-d$ | 5b | $t$ | 8d | | $0$ | $-d+t$ |
| | FLR | | $-d$ | | $-t$ | | $0$ | $0$ | $-d-t$ |
| $v_{39}$ | LDR | 9, 10c | $-d$ | 5d | $0$ | 8d | | $0$ | $-d$ |
| | FLR | | $w+d-ch$ | | $0$ | | $0$ | $c$ | $w-c+d-ch$ |
| $v_{40}$ | LDR | 9, 10b | $-d$ | 5c | $-t-b$ | 8d | | $0$ | $-d-t-b$ |
| | FLR | | $-d$ | | $t+b$ | | $0$ | $0$ | $-d+t+b$ |
| $v_{41}$ | LDR | 8a or 9, 10b | $-d$ | 5d | $0$ | 8d | | $0$ | $-d$ |
| | FLR | | $-d$ | | $0$ | | $0$ | $0$ | $-d$ |

Table 6: Payoff analysis of Game 4

Now going up, at information set $\bar{I}_{2,2} = \{v_2\}$, it is FLR's turn to move. Its expected payoff:

$$\begin{aligned} u_2(s; \bar{I}_{2,2}, \beta) &= \beta_2(v_2)u_2(s; v_2) \\ &= u_2(s; v_2) \\ &= \psi_3 u_2(v_{19}) + \psi_4 u_2(v_{28}) + \psi_5 u_2(v_{37}) \end{aligned}$$

Because $u_2(v_{28}) > u_2(v_{19}) > u_2(v_{37})$, FLR must choose $\psi_3 = 0, \psi_4 = 1, \psi_5 = 0$ to maximize its expected payoff.

Going up, at at information set $\mathcal{I}_{2,1} = \{v_1\}$, it is FLR's turn to move. Its expected payoff:

$$\begin{aligned} u_2(s; \mathcal{I}_{2,1}, \beta) &= \beta_2(v_1)u_2(s; v_1) \\ &= u_2(s; v_1) \\ &= \psi_1 u_2(\text{Game 2}) + \psi_2 u_2(v_{28}) \end{aligned}$$

Because $u_2(v_{28}) > u_2(\text{Game 2})$, FLR must choose $\psi_1 = 0, \psi_2 = 1$ to maximize its expected payoff.

At the top, it is LDR's information set $\mathcal{I}_{1,1} = \{v_0\}$. LDR's expected payoff:

$$\begin{aligned} u_2(s; \mathcal{I}_{2,1}, \beta) &= \beta_1(v_0)u_1(s; v_0) \\ &= u_2(s; v_0) \\ &= \phi_1 u_1(\text{Game 2}) + \phi_2 u_1(v_{28}) \end{aligned}$$

Because $u_1(\text{Game 2}) > u_1(v_{28})$, LDR must choose $\phi_1 = 1, \phi_2 = 0$ to maximize its expected payoff.

Consistent can be proven by using the following sequence $s^k = (s_1^k, s_2^k)$ where:

$$\begin{cases} s_1^k = & \left(\left[\frac{k-1}{k}(\neg init), \frac{1}{k}(init)\right], \left[\frac{1}{k}(f(x)), \frac{k-2}{k}(r), \frac{1}{k}(other)\right]\right) \\ s_2^k = & \left(\left[\frac{1}{k}(\neg collude), \frac{k-1}{k}(collude)\right], \right. \\ & \left[\frac{1}{k}(\neg report), \frac{k-2}{k}(report, y' = f(x)), \frac{1}{k}(report, y' \neq f(x)), \right. \\ & \left.\frac{1}{k}(f(x)), \frac{k-2}{k}(r), \frac{1}{k}(other), \frac{1}{k}(f(x)), \frac{k-2}{k}(r), \frac{1}{k}(other), \right. \\ & \left.\left.\frac{1}{k}(f(x)), \frac{k-2}{k}(r), \frac{1}{k}(other)\right]\right) \end{cases}$$

It is clear that $s^k$ is fully mixed, i.e. every pure strategy has a non-zero probability. Because $\lim_{k \to \infty} \frac{k-1}{k} = 1$, $\lim_{k \to \infty} \frac{k-2}{k} = 1$ and $\lim_{k \to \infty} \frac{1}{k} = 0$, $s^k$ converges to $s$. The induced belief system $\beta^k = (\beta_1^k, \beta_2^k)$ is:

$$\begin{cases} \beta_1^k = & \left([1(v_0)], \left[\frac{1}{k}(v_3), \frac{k-2}{k}(v_4), \frac{1}{k}(v_5)\right]\right) \\ \beta_2^k = & \left([1(v_1)], 1(v_2)], \left[\frac{1}{k}(v_6), \frac{k-2}{k}(v_7), \frac{1}{k}(v_8)\right], \right. \\ & \left[\frac{1}{k}(v_9), \frac{k-2}{k}(v_{10}), \frac{1}{k}(v_{11})\right] \\ & \left.\left[\frac{1}{k}(v_{12}), \frac{k-2}{k}(v_{13}), \frac{1}{k}(v_{14})\right]\right) \end{cases}$$

which also converges to $\beta$. □

**Proof of Theorem 7.4**

PROOF. In Lemma 7.3, we showed that the game in Figure 5 has only one sequential equilibrium. In the equilibrium LDR will play $\neg init$ with a probability 1, thus the game will always go to the Game 2 branch. In this branch, the two clouds play the misreporting game and we have showed in Theorem 7.2, the misreporting game will always terminate at $v_{13}$ in Figure 4. □

## D PSEUDOCODE
### D.1 Prisoner's Contract

**Prisoner's Smart Contract**

**Init:** Set state := INIT, deposit := {}, worker := {}, result := {}

**Create:** Upon receiving from a client CLT
  ("create", $com_f, com_x, w, d, ch, T_1, T_2, T_3$, TTP):
  Assert state = INIT and $T < T_1 < T_2 < T_3$ and
    ledger[CLT] ≥ $(2 \cdot w + ch)$
  ledger[CLT] := ledger[CLT] $-\$(2 \cdot w + ch)$
  deposit := deposit $\cup$ (CLT, $(2 \cdot w + ch)$)
  state := CREATED

**Bid:** Upon receiving ("bid") from a Cloud $C_i$:
  Assert state = CREATED and $T < T_1$ and
    $(C_i, \$d) \notin$ deposit and ledger[$C_i$] ≥ $d$
  ledger[$C_i$] := ledger[$C_i$] $-\$d$
  deposit := deposit $\cup (C_i, \$d)$
  worker := worker $\cup$ $C_i$
  if |worker| = 2 then state := COMPUTE

**Deliver:** Upon receiving ("output", $com_{y_i}$) from a cloud $C_i$:
  Assert state = COMPUTE and $T < T_2$ and
    $C_i \in$ worker and $(C_i, *) \notin$ result
  result := result $\cup (C_i, com_{y_i})$
  if |result| = 2 then state := PAY

**Pay:** Upon receiving ("pay", NIZK) from CLT:
  Assert state = PAY and $T < T_3$
  if |result| = 0 then
    ledger[CLT] := ledger[CLT] $+\$(2 \cdot w + 2 \cdot d + ch)$
    state := DONE
  else if |result| = 2 and
    $verify(NIZK, com_{y_1}, com_{y_2}) \to y_1 = y_2$ then
    ledger[$C_1$] := ledger[$C_1$] $+\$w + \$d$
    ledger[$C_2$] := ledger[$C_2$] $+\$w + \$d$
    ledger[CLT] := ledger[CLT] $+\$ch$
    state := DONE
  else state := ERROR

**Dispute:** Upon receiving ("resolve", $com_{y_t}$, $NIZK_1, NIZK_2$)
  from TTP:
  Let result = $(C_1, com_{y_1}), (C_2, com_{y_2})$
  Cheated := [false,false]
  for $i = 1$ to 2
    if $NIZK_i = NULL$ then
      Cheated[i] := true
    Else if $verify(NIZK_i, com_{y_i}, com_{y_t}) \to y_i \neq y_t$
      then Cheated[i] := true
  if Cheated[1] and Cheated[2] then
    ledger[CLT] := ledger[CLT] $+\$2 \cdot (w + d)$
  else if ¬Cheated[1] and ¬Cheated[2] then
    ledger[$C_1$] := ledger[$C_1$] $+\$w + \$d$
    ledger[$C_2$] := ledger[$C_2$] $+\$w + \$d$
  else if ¬Cheated[1] and Cheated[2] then
    ledger[$C_1$] := ledger[$C_1$] $+\$(w + 2 \cdot d - ch)$
    ledger[CLT]:= ledger[CLT] + $w + ch$
  else if Cheated[1] and ¬Cheated[2] then
    ledger[$C_2$] := ledger[$C_2$] $+\$(w + 2 \cdot d - ch)$
    ledger[CLT]:= ledger[CLT] + $w + ch$
  ledger[TTP] := ledger[TTP] $+\$ch$
  state := DONE

**Timer:** if $T \geq T_1$ and state = CREATED then
    refund(deposit)
    state := ABORTED
  else if $T \geq T_2$ and state = COMPUTE then
    state := PAY
  else if $T \geq T_3$ and state = PAY then
    for each (a,b) in result
      deposit := deposit $-(CLT, \$w) - (a, \$d)$
      ledger[a] := ledger[a] $+\$w + \$d$
    Let *res* be any amount left in deposit

ledger[CLT] := ledger[CLT] +$res
state := DONE

### Prisoner's Contract Protocol

**Init:** CTP := $G(Prisoner's Contract)$

**Outsource:** A client CLT chooses a function $f$ and input $x$ to the function, computes the commitments $com_f = com_{s_f}(f)$ and $com_x = com_{s_x}(x)$, sends the message
("create", $com_f, com_x, w, d, ch, T_1, T_2, T_3$, TTP) to CTP.
The client sends $(f, s_f, x, s_x)$ to $C_1$ and $C_2$, who verifies the values against the commitments on the blockchain. The clouds abort if the verification fails. Otherwise, each $C_i$ sends the message ("bid") to CTP before CTP.$T_1$.

**Compute:** Each $C_i$ computes $y_i = f(x)$ and also its commitment $com_{y_i} = com_{s_{y_i}}(y_i)$. It sends ("output", $com_{y_i}$) to CTP and $(y_i, s_{y_i})$ to CLT before CTP.$T_2$.

**Check:** If CTP.state = PAY and CTP.result = {}, CLT sends the message ("pay",NULL) to CTP.
Else if CTP.state = PAY and both $(y_i, s_{y_i})$ opens the commitments in CTP.result and $y_1 = y_2$, CLT generates $NIZK$ that proves $y_1 = y_2$, and sends ("pay", $NIZK$) to CTP.
Else CLT enter **Arbitration**.

**Arbitration:** CLT sends $(f(), s_f, x, s_x)$ and $(y_i, s_{y_i})$ to TTP. TTP verifies $(f(), s_f, x, s_x), (y_i, s_{y_i})$ against the commitments in CTP, aborts if verification fails.
TTP computes $y_t = f(x)$ and $com_{y_t} = com_{s_{y_t}}(y_t)$. TTP computes $NIZK_i$ to prove $y_t \stackrel{?}{=} y_i$. If $com_{y_i}$ is missing from CTP.result or $(y_i, s_{y_i})$ cannot open $com_{y_i}, NIZK_i = NULL$.
TTP sends ("resolve", $com_{y_t}, NIZK_1, NIZK_2$) to CTP.
TTP sends $(y_t, s_{y_t})$ to CLT.

### D.2 Colluder's Contract

#### Colluder's Smart Contract

**Init:** Set state := INIT, deposit := {}

**Create:** Upon receiving the message ("create", CTP, $C_2$, $com(r)_1$, $com(r)_2, t, b, T_4, T_5$) from $C_1$:
Assert state = INIT and CTP = $G(Prisoner's Contract)$
and $T < T_4 <$ CTP.$T_2 <$ CTP.$T_3 < T_5$ and
CTP.state = COMPUTE and ledger[$C_1$] $\geq$ \$$(t + b)$
ledger[$C_1$] := ledger[$C_1$] − \$$(t + b)$
deposit := deposit $\cup(C_1, \$(t + b))$
state := CREATED

**Join:** Upon receiving the message ("join") from $C_2$:
Assert state = CREATED and $T < T_4$ and
CTP.state = COMPUTE and ledger[$C_2$] $\geq$ \$$t$
ledger[$C_2$] := ledger[$C_2$] − \$$t$
deposit := deposit $\cup(C_2, \$t)$
state := COLLUDED

**Enforce:** If $T \geq T_5$ and state = COLLUDED and
CTP.state = DONE then
Let $(C_1, com_{y_1}), (C_2, com_{y_2}) =$ CTP.result
if $com_{y_1} = com_{r,1}$ and $com_{y_2} = com_{r,2}$, then
ledger[$C_1$] := ledger[$C_1$] + \$$t$
ledger[$C_2$] := ledger[$C_2$] + \$$(t + b)$
if $com_{y_1} = com_{r,1}$ and $com_{y_2} \neq com_{r,2}$, then
ledger[$C_1$] := ledger[$C_1$] + \$$(2 \cdot t + b)$

else if $com_{y_1} \neq com_{r,1}$ and $com_{y_2} = com_{r,2}$, then
ledger[$C_2$] := ledger[$C_2$] + \$$(2 \cdot t + b)$
else refund(deposit)
state:= DONE

**Timer:** If $T \geq T_4$ and state = CREATED, then
refund(deposit)
state = ABORTED

#### Colluder's Contract Protocol

The protocol is run by $C_1$ and $C_2$ during the compute step of the Prisoner's Contract protocol, before they send their messages to CLT and CTP.

**Init:** CTP := $G(Prisoner's Contract)$,
CTC := $G(Colluder's Contract)$

**Attempt:** $C_1$ chooses $r$ randomly as the computation results, computes two commitments of $r$: $com_{r,1} = com_{s_1}(r)$, $com_{r,2} = com_{s_2}(r)$. $C_1$ sends the message ("create", CTP, $C_2, com_{r,1}, com_{r,2}, t, b, T_4$) to CTC, and the message (CTC, $\$t, b, r, s_1, s_2$) to $C_2$.

**Agree:** If $C_2$ agrees to collude, it verifies that CTC.state = CREATED and the two commitments opens to $r$. It sends the message ("join") to CTC.
$C_1$ and $C_2$ will not execute the Compute step in the Prisoner's Contract Protocol, instead, they do the following: Each $C_i$ sends ("output", $(com_{r,i})$) to CTP and $(r, s_i)$ to CLT before CTP.$T_2$.

**Enforce:** After $T_5$ and CTP has concluded, any one of the clouds can send ("enforce") to CTC to enforce the agreement.

### D.3 Traitor's Contract

#### Traitor's Smart Contract
(assuming $C_2$ is the traitor)

**Init:** Set state := INIT, deposit := {}

**Create:** Upon receiving the message ("create", CTP, CTC, $C_2$) from CLT:
Assert state = INIT and CTP = $G(Prisoner's Contract)$
CTC = $G(Colluder's Contract)$ and $T <$ CTP.$T_2$
and CTC.state = CREATED or COLLUDED
Let $d, w, ch$ be the same as in CTP
Assert ledger[CLT] $\geq$ \$$(w + 2 \cdot d - ch)$
ledger[CLT] := ledger[CLT] − \$$(w + 2 \cdot d - ch)$
deposit := deposit $\cup(CLT, \$(w + 2 \cdot d - ch))$
state := CREATED

**Join:** Upon receiving the message ("join") from $C_2$:
Assert state = CREATED and ledger[$C_2$] $\geq$ \$$ch$
and CTP.state = COMPUTE and $T <$ CTP.$T_2$
ledger[$C_2$] := ledger[$C_2$] − \$$ch$
deposit := deposit $\cup(C_2, \$ch)$
state := JOINED

**Deliver:** Upon receiving the message ("output", $com_{y'}$) from $C_2$:
Assert state = JOINED and $T <$ CTP.$T_2$
and CTP.state = COMPUTE
state := COMPUTED

**Check:** Upon receiving the message ("check", $NIZK$) from CLT:
Assert state = COMPUTED and CTP.state = DONE
Cheated := CTP.dispute.Cheated

$com_{y_t} := CTP.dispute.com_{y_t}$
Correct := false
if $verify(NIZK, com_{y'}, com_{y_t}) \rightarrow y' = y_t$ then
  Correct := true
if ¬Cheated[1] and ¬Cheated[2] then
  ledger[CLT] := ledger[CLT] + $(w + 2 \cdot d)$
else if ¬Cheated[1] and Cheated[2] and Correct then
  ledger[$C_2$]:= ledger[$C_2$] + $w$
  ledger[CLT] := ledger[CLT] +$2 \cdot d$
else if Cheated[1] and Cheated[2] and Correct then
  ledger[$C_2$]:= ledger[$C_2$] + $w + 2 \cdot d$
else refund(deposit)
  state := DONE

**Timer:** If $T \geq CTP.T_2$ and state = CREATED then
  refund(deposit)
  state := ABORTED
Else if $T \geq CTP.T_2$ and state = COMPUTE then
  ledger[CLT] := ledger[CLT] + $(w + 2 \cdot d)$
  state := DONE
If $T \geq CTP.T_3$ and state = COMPUTED then
  ledger[$C_2$]:= ledger[$C_2$] + $w + 2 \cdot d$
  state := DONE

### Traitor's Contract Protocol

The protocol is run by CLT and $C_2$. The first 3 steps (Init - Join) must be run during the Agree step of the Colluder's Contract protocol, before $C_2$ sends its message to CTC. The Deliver step must be run before the Check step in the Prisoner's Contract protocol. The Check step must be run after the Prisoner's Contract protocol concluded.

**Init:** $CTP := G(Prisoner'sContract)$,
  $CTC := G(Colluder'sContract)$
  $CTT := G(Traitor'sContract)$

**Report:** Upon receiving $C_1$'s request to collude, $C_2$ verifies CTC.state = CREATED. It then sends the message (CTC, $C_2$) to CLT to report the collusion attempt. Then CLT sends ("create", CTP, CTC, $C_2$) to CTT. CLT notifies $C_2$ once the contract is created. $C_2$ delays joining the collusion until it verifies that CTT.state = CREATED. It then sends ("join") to CTT to join the Traitor's Contract. It runs the rest of Agree step in the Colluder's Contract Protocol and also computes $f(x)$. Once $C_2$ computed $f(x)$, it sets $y' = f(x)$ and computes $com_{y'} = com_{s'}(y')$. It sends ("output", $com_{y'}$) to CTT. $C_2$ also sends $(y', s')$ to CLT.

**Check:** CLT skips the Check step in the Prisoner's Contract Protocol and always enters the Arbitration step.
CLT wait until the Arbitration step finish,es then computes NIZK to prove $y_t \stackrel{?}{=} y'$ where $y_t$ was the computation result received from TTP at the end of the Arbitration step. Then it sends ("check", $NIZK$) to CTT.

## E  COMMITMENT AND NIZK

### Pedersen Commitment
- Public Parameters: $(G, P, Q)$ such that $G$ is an order-$q$ elliptic curve group over $\mathbb{F}_p$, $P$ and $Q$ are random generators of $G$.
- Commit: To commit $m \in Z_q$, choose $s \in_R Z_q$, and compute $Com_s(m) = mP + sQ$.
- Open: Given $(m, s)$ and a commitment $C$, accept only if $C = mP + sQ$.

### Equality NIZK

- Public parameters: $(G, P, Q, H)$ such that $G$ is an order-$q$ elliptic curve group over $\mathbb{F}_p$, $P$ and $Q$ are random generators of $G$ and $H : \{0, 1\}^* \rightarrow Z_q$ is a collision resistant hash function.
- To prove: For two Pederson commitments $C_1 = \alpha_1 P + \beta_1 Q, C_2 = \alpha_2 P + \beta_2 Q$, the prover knows $\alpha_1, \alpha_2, \beta_1, \beta_2$, and $\alpha_1 = \alpha_2$.
  (1) The prover chooses $\gamma \in_R Z_q$, computes the following: $t = \gamma Q$, $\delta = H(P, Q, C_1, C_2, t)$, and $\eta = (\beta_1 - \beta_2)\delta + \gamma$ and sends $\sigma_= = (t, \eta)$ to the verifier
  (2) The verifier computes $\delta = H(P, Q, C_1, C_2, t)$, checks $\eta Q \stackrel{?}{=} \delta(C_1 - C_2) + t$. Output 1 if true, otuput 0 otherwise.

### Inequality NIZK

- Public parameters: $(G, P, Q, H)$ such that $G$ is an order-$q$ elliptic curve group over $\mathbb{F}_p$, $P$ and $Q$ are random generators of $G$ and $H : \{0, 1\}^* \rightarrow Z_q$ is a collision resistant hash function.
- To prove: For two Pederson commitments $C_1 = \alpha_1 P + \beta_1 Q, C_2 = \alpha_2 P + \beta_2 Q$, the prover knows $\alpha_1, \alpha_2, \beta_1, \beta_2$, and $\alpha_1 \neq \alpha_2$.
  (1) The prover chooses $\gamma_1, \gamma_2 \in_R Z_q$, computes the following: $t_1 = \gamma_1 P, t_2 = \gamma_2 Q$, $\delta = H(P, Q, C_1, C_2, t_1, t_2)$ and $\eta_1 = (\alpha_1 - \alpha_2)\delta + \gamma_1$, $\eta_2 = (\beta_1 - \beta_2)\delta + \gamma_2$ and sends $\sigma_{\neq} = (t_1, t_2, \eta_1, \eta_2)$ to the verifier
  (2) The verifier computes $\delta = H(P, Q, C_1, C_2, t_1, t_2)$, and output 1 if both of the following two conditions are true, output 0 otherwise:
    – $\eta_1 P + \eta_2 Q = \delta(C_1 - C_2) + t_1 + t_2$ and
    – $\eta_2 Q \neq \delta(C_1 - C_2) + t_2$

## F  MORE ON THE EXPERIMENT

### F.1  Addresses

Our experiment was carried out on the official Etheruem network. Below are the account addresses and addresses of the transactions that are shown in Table 2. They can be viewed through public websites (e.g. https://etherscan.io/) or a suitable Ethereum client.

#### Contract Account Address

- Prisoner's Contract:
  0x09b61d58448d580c42b387334ac3fe28f2868887
- Colluder's Contract:
  0x255309e0612de2ab1812e21190b9a9b8f9a216d8
- Traitor's Contract:
  0x57b032d5a6adcc67739e8fd87a00c69bedbf7c65

#### Transaction Addresses

See Table 7. In the table **Init** is the transaction that send the contract code to the blockchain. Other transactions are the ones that invoke functions in the contracts. We tested the contracts with multiple execution with different parameters. For each function, we only record in the table one transaction that invokes it. There are some variations on gas consumption in each execution, but they are small and can be safely ignored.

| Contract | Transaction | Address |
|---|---|---|
| Prisoner's | Init | 0x96a20c45d7eea44ae19932a7ffecb29017d92b25b1dca851033cc4d0080ae4fe |
| | Create | 0x2de04ff3047d0bb2cb4470e500247a50eacdaca6f85ff134accfb7b4d022368e |
| | Bid | 0x9c5e04c64d1df0df9bdaa552cf97b87323ba330e25b3d076b4105d3c07c8a372 |
| | Deliver | 0xebc1893e285f21374e9c2c33bbbc4728f768e6d85d5552e701e56c4117b8445c |
| | Pay | 0x8d85f6ac943a19f9baf9cf1db439993939f78cf6934ae16eee1cf3b403a3d19b |
| | Dispute | 0x1dd851fb709d875d9f382b550032f20f24e29539336545b5fd733fd359f8951d |
| Colluder's | Init | 0x583902f2f5550af92ccbb32dd522d0b975f3cc7c308afa3db38879d76f99edb0 |
| | Create | 0xd04b7ca961a885626b4b2f62d4cffe25c38325751b0b5b1cbe3248b4a6121909 |
| | Join | 0xfbad90496c4f2416a9baf49f8fc1cce3de6666b1e73f4e3cb57622d032f68ee4 |
| | Enforce | 0xf06b571d4b89817d673fc4d079186525364dd744e6f291820469aff43b7134a0 |
| Traitor's | Init | 0x49cbcb5609997fb20d642782ba81c78b9504aedc37fd3aa9aca94e3374509f63 |
| | Create | 0x27f4db8f47a34f0027f5312a032891af2b377002f807c41272a653a6b74a0375 |
| | Join | 0x6365876c79025914ca0869d485757f148916a28ce12da00ae922520de0da8e99 |
| | Deliver | 0xdc4e83f68c83a198b4d7d6acb17c6573c2c0bcae09df41054db791d723d84da4 |
| | Check | 0xf333ed3c73fa28f1a964879ea8c6aa32cd612934e0d7b397ef330d9007767be1 |

Table 7: Transactions on the Ethereum network

## F.2 Reuse Contracts

As we can see in Table 2, Init transaction (sending the contract code to the blockchain) costs the most among all transactions when using a contract. This is due to the high cost of storing data on the blockchain. In many cases, the contracts can be reused. This is useful, for example, when a client has a sequence of computation tasks to outsource. Reusing a contract will reduce or destroy anonymity, however this might not be a big concern in the commercial world.

In our contracts, we have a reset function that can be called by the owner after the contract has concluded. The function will clean up all data and reset the contract to the initial state. Then the contract can be used as a new one. The cost of calling the reset function is considerably smaller than setting up a new contract. For example, resetting a Prisoner's contract only costs about 56 thousand gas ($0.01), while setting up a new contract costs about 2.3 million gas ($0.40). This could help to reduce significantly the transaction cost per contract execution.

## G IS SIGNING MESSAGES ENOUGH TO DETER CHEATING?

A question raised in the review process was that whether something simple like just having the cloud providers sign messages is enough. More specifically, one can design a mechanism in which the two clouds sign all messages, including the one that delivers the result. The client retains all the signed messages. If later the client finds the result from a cloud is not correct, it can expose the wrongdoing with the signature as evidence to the public, which will damage the reputation of the cloud and/or cause the cloud a financial loss. The clouds could be deterred by the consequences and thus would not cheat in the first place.

This simple mechanism could work if the following two assumptions hold: (1) the clouds do not collude; (2) exposing the wrongdoing will cause enough damage to the clouds and the clouds cannot avoid it. The first assumption ensures that the client can at least detect cheating (given that the client is assumed to be resource limited and cannot recompute the tasks by itself), and second ensures that the clouds have the incentive to not to cheat.

In our paper, we consider cases in which collusion is a possibility that cannot be eliminated. In real world, there are cases that the clouds prefer not to collude, e.g. if collusion incurs a high cost that exceeds the benefit of collusion. In such cases, if the client believes that the clouds will not collude then the counter-collusion contracts are not necessary because collusion is no longer a threat, and the client could adopt simpler solutions.

In our paper, the deposit is held in advance and the cheating cloud cannot get away from the penalty because the contracts are enforced automatically and faithfully, assuming the smart contract network is not compromised. In this case, the damage to the cheating cloud by losing deposit is concrete and invariable. On the other hand, the damage caused by losing reputation depends on assumptions on practical factors such as the voice of the client, and the effectiveness of the PR/legal departments of the cloud provider. Some cloud providers might be able to avoid the damage by e.g. changing their identities. If the damage is not enough or the clouds can get away from penalty, threatening with exposing wrongdoing will not deter cheating.